\documentclass[5p]{elsarticle}

\biboptions{sort&compress}

\usepackage{amsmath}
\usepackage{amssymb}
\usepackage{amsxtra,color}
\usepackage{latexsym}
\usepackage{graphicx}

\newcommand{\braOket}[3]{\langle#1 |#2|  #3 \rangle}

\begin{document}

\title{Theory of attosecond delays in laser-assisted photoionization}

\author[lth,su]{J.~M.~Dahlstr\"om}\ead{marcus.dahlstrom@fysik.su.se}
\author[lth]{D.~Gu\'enot}
\author[lth]{K.~Kl{\"u}nder} 
\author[lth]{M.~Gisselbrecht}
\author[lth]{J.~Mauritsson}
\author[lth]{A.~L'Huillier}\ead{anne.lhuillier@fysik.lth.se}
\author[upmc,cnrs]{A.~Maquet}
\author[upmc,cnrs]{R.~Ta\"ieb}\ead{richard.taieb@upmc.fr}

\address[lth]{Department of Physics, Lund University, P.O. Box 118, 22100 Lund, Sweden}
\address[su]{Atomic Physics, Fysikum, Stockholm University, AlbaNova University Center, SE-106 91 Stockholm, Sweden}
\address[upmc]{UPMC Universit\'e Paris 6, UMR 7614, Laboratoire de Chimie Physique-Mati\`ere et Rayonnement, \\
11 rue Pierre et Marie Curie, 75231 Paris Cedex 05, France}
\address[cnrs]{CNRS, UMR 7614, LCPMR, Paris, France}


\begin{abstract}
We study the temporal aspects of laser-assisted extreme ultraviolet (XUV) photoionization  
using attosecond pulses of harmonic radiation.
The aim of this paper is to establish the general form of the {\it phase} of the relevant transition amplitudes and to make the connection with the time-delays that have been recently measured in experiments. We find that the overall phase contains two distinct types of contributions: one is expressed in terms of the phase-shifts of the photoelectron continuum wavefunction while the other is linked to continuum--continuum transitions induced by the infrared (IR) laser probe. Our formalism applies to both kinds of measurements reported so far, namely the ones using attosecond pulse trains of XUV harmonics and the others based on the use of isolated attosecond pulses (streaking). 
The connection between the phases and the time-delays is established with the help of finite difference approximations to the energy derivatives of the phases. The observed time-delay is a sum of two components: 
a one-photon Wigner-like delay and an {\it universal} delay that originates from the probing process itself. 
\end{abstract}

\maketitle 

\section{Introduction}
\label{sec:intro}

The dynamics of photoionization can now be explored with unprecedented
time resolution thanks to high-order harmonic-based sources that deliver pulses of  XUV radiation with duration in the
attosecond range. Recent measurements performed with single attosecond pulses have shown the existence of
an unexpected time-delay between the single-photon ionization from the 2s and the 2p sub-shells
of Ne atoms in gas phase \cite{Schultze2010}. The ``streak camera'' technique used in these experiments \cite{Goulielmakis2004}  implied nontrivial ejection times of the photoelectrons, depending on the sub-shell from which they originate. 
Similar delays between the ejection times from the 3s and 3p sub-shells in Ar have been measured also
using trains of attosecond pulses 
 \cite{Klunder2011}, with the help of another technique based on interferometry called RABBIT (Reconstruction of Attosecond Beating By Interference of Two-photon transitions)
\cite{Veniard95,Veniard96,Muller2002}. 
In both cases, delays of several tens of attoseconds have been measured. 
As photoionization is one of the most fundamental processes in light-matter interactions, these results have motivated a large number of theoretical investigations \cite{Yakovlev2010, Kheifets2010,Ivanov2011,Zhang2011,Nagele2011, IvanovM2011,Pazourek2011}. 


\begin{figure*}[t]
	\centering
		\includegraphics [width=0.9\textwidth]{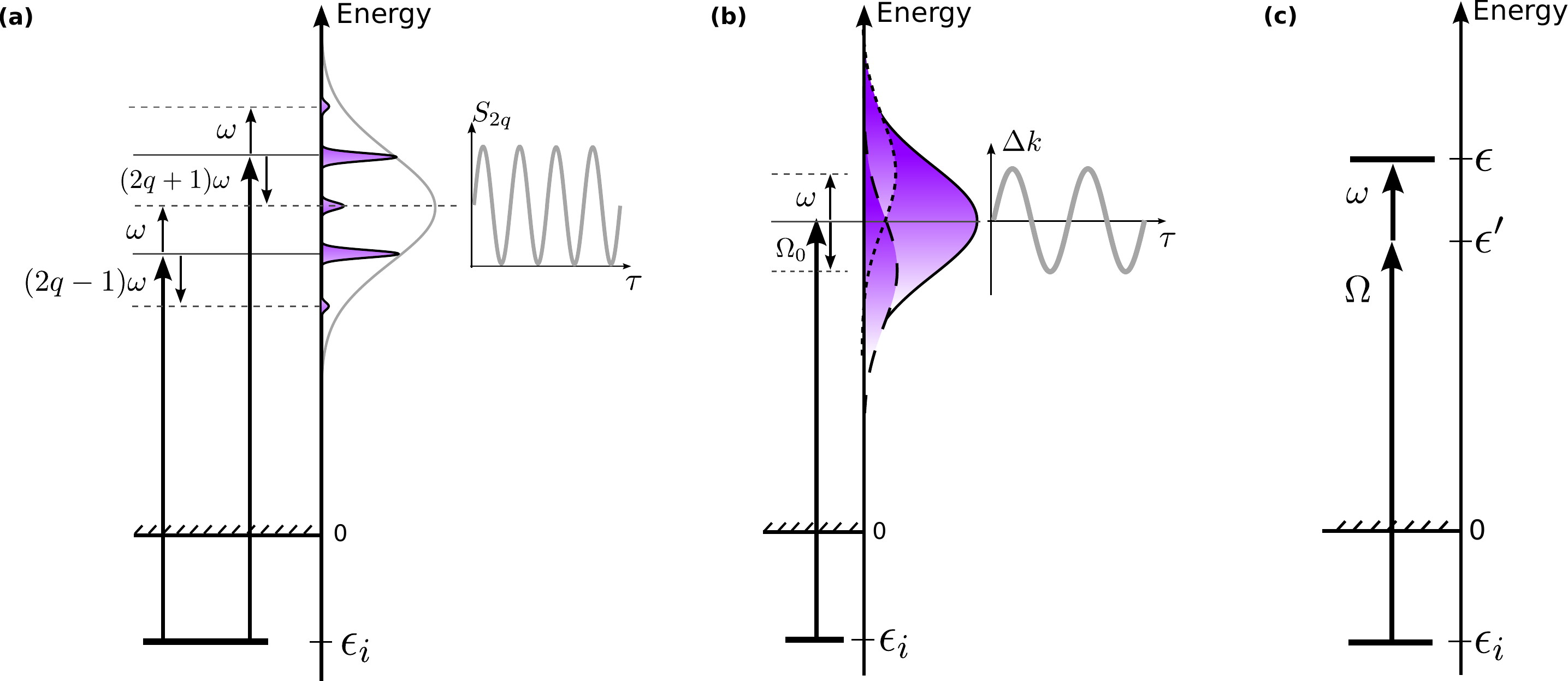}
	\caption{\label{fig1}
(a) Laser-assisted photoionization by an attosecond pulse train, 
corresponding to odd XUV harmonics, $H_{2q-1}$, where $q$ is a positive integer.
The sideband $S_{2q}$ can be reached by either 
absorbing $H_{2q-1}$ and then absorbing a IR laser photon $\omega$
or by absorbing $H_{2q+1}$ and then emitting $\omega$. 
The sideband signal, $S_{2q}$, oscillates as a function of the subcycle-delay, $\tau$, between the attosecond pulses and the IR laser probe field. 
(b) Laser-assisted photoionization by a single attosecond pulse, 
corresponding to a broad XUV continuum.
To first order, the electron is ionized by absorbing one XUV photon, $\Omega$, 
resulting in a wave packet centered at $\epsilon=\epsilon_i+\Omega_0$.
To second order, the electron may absorb an additional laser photon $\omega$ resulting in an upshifted wave packet
centered at $\epsilon=\epsilon_i+\Omega_0+\omega$;
or
it may emit a laser photon resulting in a downshifted wave packet 
centered at $\epsilon=\epsilon_i+\Omega_0-\omega$. 
The interference of these three wave packets leads to a modulation, $\Delta k$, of central momentum 
of the photoelectron as function of the subcycle-delay, $\tau$, between the laser field and the attosecond pulse. 
(c) Two-photon XUV-IR Above-Threshold Ionization from an initial bound state with energy $\epsilon_i$, 
to a final state with energy $\epsilon$. 
	}
\end{figure*}

The two kinds of measurements  share many similarities since they involve a laser-assisted single-photon ionization process and they rely on a phase-locked IR laser field to probe the temporal aspects of the XUV photoionization. However, they differ in the analysis used to determine the time-delays and in the range of IR laser intensity. 

The motivation of the present paper is to present an unified theoretical analysis of these processes. To achieve this goal, we shall  expose first the theoretical background which has conducted us to conclude in \cite{Klunder2011}, that in 
interferometric measurements, the measured delays arise from the combination 
of two distinct contributions: One is related to
the electronic  structure of the atomic target while the other is induced by the measurement process itself. 
The first one can be identified as a ``Wigner time-delay'' \cite{Wigner1955, Nussensveig2002}
that is directly related to the energy dependence of the different phase-shifts experienced by the photoelectrons ionized
from distinct sub-shells in atoms. The other contribution is induced by the IR
laser field that is used to probe the photoionization process. This latter contribution results from the continuum--continuum transitions induced by the probe IR laser field in the presence of the Coulomb potential
of the ionic core. When simplifying the analysis to the cases when the process is dominated by the asymptotic form of the relevant second-order matrix elements, a characteristic measurement-induced delay can be identified, that is independent from the details of the electronic structure of the ionic core. This shows how the experimental signal can be related to the temporal dynamics of one-photon ionization.

Regarding the streaking measurements realized with a single attosecond pulse of XUV radiation \cite{Schultze2010}, the experimental data were obtained for IR field intensities significantly higher than those obtained with attosecond pulse trains \cite{Klunder2011}. Understandably,  the questions related to the role of the probe IR field on the photoelectron dynamics in streaking measurements have motivated several theoretical studies \cite{Yakovlev2010, Kheifets2010,Ivanov2011,Zhang2011,Nagele2011, IvanovM2011,Pazourek2011}; see also the earlier papers: \cite{Smirnova2006, Smirnova2007, Zhang2010, Baggesen2010}. Then, a natural issue arises which is to determine to what extent the ``streaking delays'' so obtained differ from those derived from the interferometric data. Although both the experimental techniques and the  theory treatments differ, it is of interest to  compare the two approaches. 
Indeed, as we shall show below, a link can be found when reducing the laser intensity of the streaking field so that one reaches the domain of applicability of the recently developed Phase-Retrieval by Omega Oscillation Filtering (PROOF) scheme, \cite{Chini2010}. An interesting outcome of our analysis is to show the importance of the long-range Coulomb potential for understanding the absolute time-delays in the streaking experiments as well.

The interpretation of the attosecond delays in photoionization relies on  our ability to determine the {\em phases}  of the relevant transition amplitudes. Thus, before going into the details of the derivation of such phases, we shall outline the main features of the two techniques in Section~\ref{sec:ATI}. Then, Section~\ref{sec:ATIphases} is devoted to the presentation of the general expressions for two-color,  two-photon, complex transition amplitudes that are relevant for Above-Threshold Ionization (ATI) in single-active electron systems. The theoretical background is based on a perturbative approach and the emphasis will be on the derivation of a closed-form approximate expression that is of interest for evaluating the phase of the amplitudes. The basis of exact computations in hydrogen will be outlined, and a simplified classical treatment will be presented. 
Applications to the determination of the relation between the phases and the time-delays is presented in Section~\ref{sec:ATIapplications}. Here we consider first ionization by an attosecond pulse train and then by a single attosecond pulse, in the presence of a relatively weak IR field. This discussion provides an interesting connection between the two types of measurements. Section~\ref{sec:results} contains a comparison of the results extracted from the approximate evaluation of the delays to the ones deduced from exact calculations performed in hydrogen from different initial states. Also, we present our conclusions and perspectives.

\section{Laser-Assisted XUV Photoionization: Attosecond Pulse Train vs. Single Attosecond Pulse}
\label{sec:ATI}
The principle of the  measurements of the delays using an attosecond pulse train is illustrated in Fig.~\ref{fig1}~(a), which represents schematically the ionization of an atom in the simultaneous presence of a set of several XUV (odd) harmonics and of the IR field,  used to generate the harmonics (atomic units will be used throughout the paper, unless otherwise stated). In the time domain, both pulses are ``long'', {\it i.e.} the IR laser pulse is multi-cycle, with typical duration of a few tens of femtoseconds, and the XUV harmonic field is constituted of a train of attosecond pulses (or equivalently of a comb of coherent odd harmonic frequencies $(2q+1)\omega$: $H_{2q+1}$). Under these conditions, the photoelectron spectrum consists of equidistant lines separated by $2 \omega$ that are associated to one-photon ionization of the target by each harmonic. In-between these lines are sidebands associated to two-photon transitions involving the absorption of one harmonic and the exchange of one IR photon. The signal intensities, $S_{2q}$, of the sidebands labelled $2q$ vary periodically with the delay $\tau$ between the IR and the harmonic pulses, according to a generic expression that involves the phases of the fields together with atom-dependent contributions:
\begin{align}
S_{2q} = {} & \alpha + \beta \cos[2 \omega \tau - \Delta \phi_{2q} - \Delta \theta_{2q} ],
\label{S2q}
\end{align}
where $\Delta \phi_{2q} = (\phi_{2q + 1} -\phi_{2q-1})$ is the phase difference between the consecutive harmonics $H_{2q+1}$ and $H_{2q-1}$ and $\Delta \theta_{2q} $ is an intrinsic atomic quantity, associated to the difference of the phases of the transition amplitudes associated to the distinct quantum paths leading to the sideband~\cite{Veniard96}.  

To make clearer the connection between the above phase differences and the time-delays we shall discuss here, it is convenient to rewrite the formula in Eq.~(\ref{S2q}) under the form: \begin{align}
\label{S2qt}  
S_{2q} = {} & \alpha + \beta \cos [2\omega (\tau - \tau_{2q} - \tau_\theta) ], 
\end{align} 
where $\tau_{2q} = \Delta \phi_{2q}/2\omega$ is a finite difference approximation to the group delay $GD = \partial \phi /\partial \Omega$ of the harmonic radiation at the considered frequency, $\Omega\approx 2q \omega $, as presented in refs.~\cite{Paul2001, Mairesse2003}. On the other hand, $\tau_\theta= \Delta \theta_{2q}/2\omega$ is an intrinsic time-delay associated to the atomic phase difference, $\Delta \theta_{2q}$. 
As reported in \cite{Klunder2011} and as we shall describe in more details here, the determination of $\tau_\theta$ gives access to the temporal dynamics of atomic photoionization. Before closing this brief presentation of the RABBIT scheme, we stress that the intensities of both fields must be kept moderate, so that the phases of the transition amplitudes associated to the sidebands can be derived from a standard time-dependent perturbation theory calculation, limited to second-order.

As represented schematically in Fig.~\ref{fig1}~(b), streaking relies on the ionization of the atom by a single attosecond  pulse, in the presence of a few-cycle IR pulse. One requirement to realize streaking is that the effective duration of the attosecond pulse has to be significantly shorter than the IR pulse cycle \cite{Goulielmakis2004} (more rigorously it is actually the spectral bandwidth of the attosecond pulse that must be larger than the probe photon frequency). The measurement consists then in recording the momentum, ${\vec k}_f (t)$, of the ejected photoelectron, as deflected by the instantaneous IR probe vector potential, ${\vec A}_\omega (t)$, so that its wave vector is given approximately by:
\begin{align}
\label{Streak}  
{\vec k}_f \approx {\vec k} - {\vec A}_\omega (t),
\end{align}  
where ${\vec k}$ is the field-free momentum. We note that this simple relation is being derived by assuming that the photoelectron does not experience the effects of the residual Coulomb potential of the ionic core \cite{Goulielmakis2004}. 
In this article, we will recover this streaking phenomenon using an interferometric interpretation, 
which includes the full effect of the ionic core, thereby, obtaining the correct absolute delay of the momentum modulation relative to the probe field.


\begin{figure}[h]
	\centering
		\includegraphics [width=0.40\textwidth]{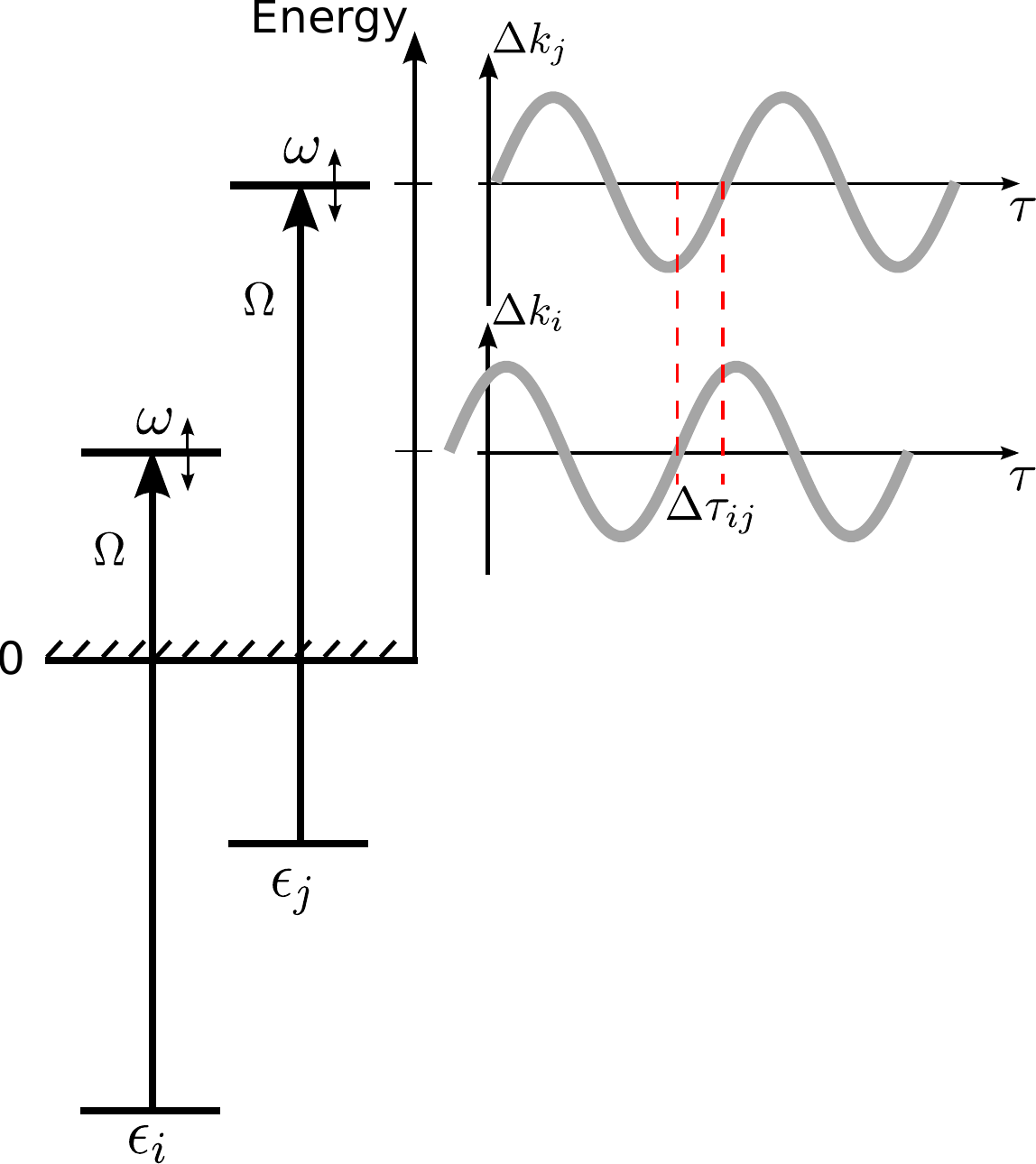}
	\caption{\label{fig2}
Sketch of an attosecond photoionization time-delay experiment 
between two different initial states with energies, $\epsilon_i$ and $\epsilon_j$.
The attosecond XUV field, $\Omega$, ionizes the electron in the presence of a phase-locked IR
laser field, $\omega$. 
Since the same single attosecond pulse (or attosecond pulse train) is used to promote the electrons from either state,
the observed delays of the modulation of the central momenta (or sidebands), $\Delta \tau_{ij}(\Omega)$, 
are directly related to the atomic delay difference $\tau_{\theta}^{(i)}(\Omega)-\tau_{\theta}^{(j)}(\Omega)$ defined in Eq.~(\ref{S2qt}), which, in turn, is related to the difference of the corresponding two-photon matrix element phases.}
\end{figure}

If the XUV field frequency is high enough to ionize electrons from either one of two valence sub-shells of an atom (typically the 2s and 2p states in Ne atoms) one can record two distinct streaking traces corresponding to the two photoelectron lines associated to each sub-shell. This situation is schematically displayed in Fig. 2. A delay between the ejection times of the photoelectrons from the different atomic sub-shells can be determined by comparing the corresponding streaking traces, corrected from possible biases introduced by the experimental procedure \cite{Schultze2010}. The same idea can be applied to attosecond pulse trains, by using the interferometric set-up \cite{Klunder2011}. 
We turn now to the presentation of the theoretical background that is common to the two kinds of techniques,
in the limit of weak IR probe fields.  


\section{Phases of Laser-Assisted Photoionization Transition Amplitudes}
\label{sec:ATIphases} 
\subsection{Two-photon Above-Threshold Ionization }

A representative laser-assisted photoionization transition is depicted in Fig.~\ref{fig1}~(c): It displays the sequential absorption of one XUV harmonic photon with frequency $\Omega$, followed by the absorption of one IR laser photon with frequency $\omega$. It corresponds to the lowest-order perturbative amplitude for an Above-Threshold Ionization (ATI) process observed when the XUV frequency is larger than the ionization energy of the system: $\Omega > I_p$. Obviously, other quantum paths are contributing to  this  type of two-color ionization process, 
{\it e.g.} the IR photon can be absorbed before the XUV photon, 
but ATI amplitudes of the former type are dominant in the class of experiments considered here. 

For two fields with the same linear polarization ${\vec \epsilon}$, it is natural to choose this direction as the quantization axis $\hat z$, and the matrix element associated to the path shown in Fig.~\ref{fig1}~(c), is of the form:
\begin{align}
M ({\vec k};\epsilon_i +\Omega) = {} &   \frac{1}{i} E_\omega E_\Omega  \lim_{ \varepsilon \to  0^+} \int \!\!\!\!\!\!\!\! \sum_{\nu}  {\langle \: \vec k \: | \: z \:  | \: \nu \: \rangle  \langle \: \nu \: | \: z \: | \: i \: \rangle\over \epsilon_i +\Omega - \epsilon_\nu  +i \varepsilon}  ,  
\end{align} 
where $E_\Omega$ and $E_\omega$ are the complex amplitudes of the harmonic  and IR laser fields, respectively; $\varphi_{n_i, \ell_i,m_i} ({\vec r})=\langle \vec r |  i \rangle$ is the initial state wavefunction, with negative energy $\epsilon_i$ and  $\varphi_{\vec k} ({\vec r})=\langle \vec r |  \vec k \rangle$ is the final state wavefunction with positive energy $\epsilon_k = {k^2 / 2} = \epsilon_i +\Omega +\omega$. The sum over the index $\nu$ runs over the whole spectrum (discrete plus continuous) of the atom. The partial wave expansion of the final state wavefunction is:
\begin{align*}
 \varphi_{\vec k}({\vec r}) =  {} &  (8 \pi)^{3/2} \sum_{L,M} i^{L} e^{ -i \eta_L (k)} Y_{L,M}^*({\hat k}) Y_{L,M}({\hat r}) R_{k,L}(r),
\end{align*} 
where the $Y_{L,M}$ are spherical harmonics, the $R_{k,L}(r)$ are (real) radial wavefunctions normalized on the energy scale and $\eta_{L} (k)$ are the phase-shifts. We note that this wavefunction behaves asymptotically as the superposition of a plane wave plus an {\em ingoing} spherical wave, as required to treat photoionization \cite{Landau}. Thus, the phase-shifts $\eta_{L} (k)$ account for the phase difference between the free motion of a plane wave and that of a photoelectron wave ejected from an atomic bound state. 

The angular dependence of the matrix element can be factorized out for an initial state 
\begin{align*}
\varphi_{n_i,\ell_i,m_i} ({\vec r })= {} &  Y_{\ell_i,m_i}({\hat r}) R_{n_i,\ell_i}(r)
\end{align*}
and with $z = \sqrt{4 \pi/3} \: r \:Y_{1,0}({\hat r})$, it becomes: 
\begin{align}
\label{Ma}
M({\vec k},\epsilon_i+\Omega) =  {} &  
\:{4 \pi \over 3i} ~ (8 \pi)^{3/2} ~ E_\omega E_\Omega ~  \nonumber \\  
\times ~ & \sum_{L,M} (-i)^{L} e^{ i \eta_{L}(k)} Y_{L,M}({\hat k})  \nonumber \\ 
\times ~ & \sum_{\lambda,\mu}  \langle Y_{L,M}| Y_{1,0}|Y_{\lambda,\mu} \rangle   \langle Y_{\lambda,\mu}| Y_{1,0}| Y_{\ell_i,m_i}\rangle \nonumber \\
\times ~ & T_{L,\lambda,\ell_i}( k, \epsilon_i +\Omega), 
\end{align} 
 where  the angular momentum components of the intermediate states are labelled $(\lambda, \mu)$ and the quantity, $T_{L,\lambda,\ell_i}(k; \epsilon_i +\Omega)$, is the radial part of the amplitude. The span of accessible angular momentum states in the intermediate and final states is governed by the dipole selection rules: One has $\lambda = \ell_i \pm 1; L = \ell_i, \ell_i\pm 2$ and $M = \mu =m_i$ respectively. The explicit form of the radial amplitude $T_{L,\lambda,\ell_i}( k;\epsilon_i +\Omega)$ is: 
\begin{align} 
T_{L,\lambda,\ell_i}(k; \epsilon_i +\Omega) 
= {}  \sum_{\nu:\epsilon_\nu<0} {\langle R_{k,L}| r |R_{\nu,\lambda}\rangle  \langle R_{\nu,\lambda}| r |R_{n_i,\ell_i} \rangle \over \epsilon_i+ \Omega - \epsilon_\nu } \notag  \\ 
+  \lim_{ \varepsilon \to  0^+} \int_0^{+\infty} d\epsilon_{\kappa'} {\langle R_{k,L}| r |R_{\kappa',\lambda} \rangle  \langle R_{\kappa',\lambda}|   r |R_{n_i,\ell_i} \rangle \over \epsilon_i +\Omega - \epsilon_{\kappa'}  +i \varepsilon}, 
\label{Tell}    
\end{align} 
where we have separated the contributions of the discrete and continuous spectra.

Since the frequency of the XUV harmonic is larger than the ionization potential of the atom,  $\Omega > I_p = |\epsilon_i|$, it is also larger than the excitation energies of the atom, $\Omega > \epsilon_\nu - \epsilon_i$. Accordingly, the denominators of the terms in the sum over the discrete states are positive and relatively large, which makes the overall contribution of these terms significantly smaller than that of the continuous spectrum. In the integral running on the continuous spectrum of energies $\epsilon_{\kappa'} = \kappa'^2 /2$, the denominator becomes zero at the energy  $\epsilon_\kappa = \kappa^2 /2 = \epsilon_i +\Omega$. Taking the limit $\varepsilon \to 0^+$, the integral becomes:
\begin{align}  
\lim_{ \varepsilon \to  0^+} \int_0^{+\infty} d\epsilon_{\kappa'} {\langle R_{k,L}| r |R_{\kappa',\lambda} \rangle  \langle R_{\kappa',\lambda}|   r |R_{n_i,\ell_i} \rangle \over \epsilon_i +\Omega - \epsilon_{\kappa'}  +i \varepsilon} \nonumber \\ 
= {\mathcal P} \int_0^{+\infty} d\epsilon_{\kappa'} {\langle R_{k,L}| r |R_{\kappa',\lambda} \rangle  \langle R_{\kappa',\lambda}|   r |R_{n_i,\ell_i} \rangle \over \epsilon_i +\Omega - \epsilon_{\kappa'} } \nonumber \\
 -i\pi \langle R_{k,L}| r |R_{\kappa,\lambda} \rangle  \langle R_{\kappa,\lambda}|   r |R_{n_i,\ell_i} \rangle.
     \label{Cauchy}
 \end{align} 
where the first term is the Cauchy principal value of the integral, which turns out to be real, and the second term is purely imaginary. The latter is associated with a two-step transition as it contains the product of the one-photon ionization amplitude towards the state  of energy $\epsilon_{\kappa} = \epsilon_i + \Omega$, times the continuum--continuum transition amplitude from $\epsilon_{\kappa}$ towards the final state of energy $k^2 /2 = \epsilon_{\kappa} + \omega$ that is reached upon the absorption of the IR photon $\omega$. 

The overall phase of the radial matrix element, Eq.~(\ref{Tell}), is thus governed by the ratio of the imaginary term in Eq.~(\ref{Cauchy}) to the sum of the integral principal part in the same equation plus the contribution of the discrete spectrum contained in Eq.~(\ref{Tell}). 
Accurate computations of such amplitudes and phases represent a formidable task for most atomic systems. 
This entails to rely on approximate representations of the atomic potential for each angular momentum dependent state \cite{Toma,Kennedy}
or to use many-electron techniques \cite{LHuillier1986,KheifetsComment}. There is however the notable exception of  hydrogenic systems, where ``exact'' calculations of these amplitudes are feasible \cite{Zernik, Klarsfeld79a, Klarsfeld79b,Aymar}, see below. It is thus of importance to derive an approximate treatment which should allow to get correct estimates of the phases of interest to address the questions of the time-delays. 

\subsection{Asymptotic approximation for 2-photon ATI matrix elements}
\label{sec:asympapp}

Let us re-express the radial amplitude in terms of the first-order {\em perturbed wavefunction} denoted $\rho_{\kappa,\lambda}(r)$: 
\begin{align}
\label{Trho}
T_{L,\lambda,\ell_i}(k; \epsilon_i +\Omega)  =
\braOket{R_{k,L}}{r}{\rho_{\kappa,\lambda}}.
\end{align}
 The function $\rho_{\kappa,\lambda}(r)$ solves the inhomogeneous differential equation:
\begin{align}
\label{eqdiffrho}
[H_\lambda - \epsilon_{\kappa}]\rho_{\kappa,\lambda}(r) = 
- r R_{n_i,\ell_i} (r),
\end{align}
where $H_\lambda$ is the radial atomic Hamiltonian for angular momentum $\lambda$. The fully developed eigenfunction expansion of $\rho_{\kappa,\lambda}(r)$ can be identified using Eqs.~(\ref{Tell}) and (\ref{Cauchy}):
\begin{align}
\label{rho}
\rho_{\kappa,\lambda}(r) ={}&
 \sum_{\nu:\epsilon_\nu<0} {R_{\nu,\lambda}(r)  \langle R_{\nu,\lambda}| r |R_{n_i,\ell_i} \rangle \over  \epsilon_{\kappa} - \epsilon_\nu } 
\nonumber \\
+~& {\mathcal P} \int_0^{+\infty} d\epsilon_{\kappa'} {R_{\kappa',\lambda}(r)  \langle R_{\kappa',\lambda}|   r |R_{n_i,\ell_i} \rangle \over \epsilon_{\kappa} - \epsilon_{\kappa'} } 
\nonumber \\
-~&i\pi  R_{\kappa,\lambda}(r)  \langle R_{\kappa,\lambda}|   r |R_{n_i,\ell_i} \rangle.
\end{align}
We note that it describes the radial part of the \textit{intermediate} photoelectron wave packet created upon absorption 
of the XUV photon ${\Omega}$, before absorbing the IR laser photon $\omega$.

The essence of the approximate treatment that we have implemented to get an estimate of $T_{L,\lambda,\ell_i}(k; \epsilon_i +\Omega) $, is based on using the asymptotic forms of both the final state function $R_{k,L}(r)$ and of the perturbed wavefunction $\rho_{\kappa,\lambda}(r)$ for large values of their radial coordinate. This is {\it a priori} justified by the fact that we are interested in the {\it phases} of the amplitudes which are governed by the asymptotic behavior of these functions. As an additional verification, we will compare the predictions of the model to those derived from an exact treatment in Hydrogen.

The asymptotic limit of the radial continuum wavefunction of the final state with angular momentum $L$ is of the generic form
\cite{Landau}:
\begin{align}
\lim_{r\rightarrow \infty} R_{k,L}(r) \ = \  {N_{k} \over r} \sin[k r+\Phi_{k,L}(r)],  
\label{RkL}
\end{align} 
 where $N_{k} = \sqrt{2/(\pi k)}$ is  the normalization constant in the energy scale and the phase has the general form: 
\begin{align}
 \Phi_{k,L}(r) =  Z \ln(2 k r)/k +\eta_L(k)  - \pi L /2.
 \label{Phase}
\end{align}
We note that this phase includes the logarithmic divergence 
characteristic of the Coulomb potential of  the ionic core with charge $Z$, in the asymptotic region.  The  Coulomb potential influences also the scattering phase-shift $\eta_L (k)$, which can be rewritten under the form: $\eta_L (k)  = \sigma_L (k)+\delta_L (k)$ where $\sigma_L=\arg[\Gamma(L+1-iZ/k)]$ is the Coulomb phase-shift and
where the correction $\delta_L (k)$ originates from the short range deviation of the ionic potential from a pure Coulomb potential, see for instance \cite{Kennedy,KheifetsComment}. Obviously, in the case of an hydrogenic system, one has $\delta_L (k) =0$.

To derive the asymptotic form of the perturbed wavefunction $\rho_{\kappa,\lambda}(r)$, it is in principle enough to establish the limiting structure of the differential equation it verifies. From Eq.~(\ref{eqdiffrho}), one observes that, in the asymptotic limit $r \to \infty$, the second member vanishes, as a result of the exponential decay of the bound state wavefunction $R_{n_i,\ell_i}(r)$. One is left with a standard Schr\"odinger equation for positive energy $\epsilon_{\kappa}$ which is  solved by imposing outgoing wave boundary conditions to the solutions \cite{Aymar}:  
\begin{align}
\lim_{r\rightarrow\infty} \rho_{\kappa,\lambda}(r) \propto
  {N_{\kappa}\over r}\ \exp\left[ i(\kappa r+\Phi_{\kappa,\lambda}(r))
\right] . 
\label{outgoing}
\end{align}
It is also of interest to derive explicitly the limiting forms of the terms entering the expression of $\rho_{\kappa,\lambda}(r)$ in Eq.~(\ref{rho}). 
Regarding its real part, the sum over the discrete states $\nu$ can be neglected, as each term goes asymptotically to zero. Thus, for large $r$, it reduces to a principal part integral: 
\begin{align}
\label{Rerho1}
\Re[\rho_{\kappa,\lambda}(r)] \approx 
{\mathcal P} \int_0^{+\infty} d\epsilon_{\kappa'} {R_{\kappa',\lambda}(r)  \langle R_{\kappa',\lambda}|   r |R_{n_i,\ell_i} \rangle \over \epsilon_{\kappa} - \epsilon_{\kappa'} }. 
\end{align}
This integral can be estimated by extending the integration range $\epsilon_{\kappa'} \to -\infty$ and replacing the radial continuum function $R_{\kappa',\lambda}(r)$ by its asymptotic limit according to the prescription in Eq.~(\ref{RkL}). Then, writing the sine function under its exponential form and performing  contour integrations, with semi-circles around the pole at $ \epsilon_{\kappa}$, one gets:
\begin{align}
\lim_{r\rightarrow\infty} \Re[\rho_{\kappa,\lambda}(r)] \approx -  {\pi N_{\kappa}\over r}
				\cos[\kappa r + \Phi_{\kappa,\lambda}(r)] \braOket{R_{\kappa,\lambda}}{r}{R_{n_i,\ell_i}}.
	\label{rerho}
\end{align}

Regarding the imaginary part given by the last term in Eq.~(\ref{rho}), it is enough to substitute again the asymptotic form of $R_{\kappa,\lambda}(r)$, so that:
\begin{align}
\lim_{r\rightarrow\infty} \Im[\rho_{\kappa,\lambda}(r)] \approx -  {\pi  N_{\kappa} \over r}
				\sin[\kappa r + \Phi_{\kappa,\lambda}(r)] \braOket{R_{\kappa,\lambda}}{r}{R_{n_i,\ell_i}}.
	\label{imrho}
\end{align} 
Then by regrouping the real and imaginary parts, one gets the final expression:
\begin{align}
\lim_{r\rightarrow\infty} [\rho_{\kappa,\lambda}(r)] \approx -  {\pi  N_{\kappa} \over r}
				\exp\left[
i\kappa r + i\Phi_{\kappa,\lambda}(r)\right] \braOket{R_{\kappa,\lambda}}{r}{R_{n_i,\ell_i}},
	\label{asymprho}
\end{align} 
which corresponds to a complex outgoing wave \cite{Aymar}, as expected from Eq.~(\ref{outgoing}), weighted by the dipole matrix element associated to the one-photon transition from the initial state. We note that adopting the so-called ``pole-approximation'',
which consists in neglecting the off-shell part ({\it i.e}. the real part given in Eq.~(\ref{rerho})), would lead 
to a loss of the phase information of the process since the perturbed wavefunction then would be a standing wave rather than an outgoing wave. 

The corresponding asymptotic approximation for the second-order radial matrix element, Eq.~(\ref{Tell}), is obtained by substituting in Eq.~(\ref{Trho}) the asymptotic expressions, Eqs.~(\ref{RkL}) and (\ref{asymprho}), for the radial wavefunctions of the final state and of the intermediate state, respectively.  One has explicitly: 
\begin{align}
T_{L,\lambda,\ell}(k;\epsilon_{\kappa})  \approx {} & 
- \pi \ \braOket{R_{\kappa,\lambda}}{r}{R_{n_i,\ell_i}}N_{k} N_{\kappa}  \nonumber \\ 
\times~&\int_0^{\infty} dr 
		\sin[kr+\Phi_{k,L}(r)] \  r  \ e^{\left[ i(\kappa r+\Phi_{\kappa ,\lambda}(r)) \right] }.\
		\label{Tas1}
\end{align}
To introduce the next step in our approximate treatment, one rewrites the sine in its exponential form and develop the expressions of the phases $\Phi_{\kappa,\lambda}(r)$ as given in Eq.~(\ref{Phase}). One is left with two distinct contributions containing integrals either of the type $J_+$ or $J_-$ that are defined as follows:
\begin{align}
J_\pm 	= {}&  \pm\frac{1}{2i} \int_0^{\infty} dr ~ r^{1+ iZ(1/ \kappa \pm 1/ k )} \exp\left[i(\kappa \pm k)r\right] 
\nonumber \\ 
	= {}&  \pm\frac{1}{2i} \left( \frac{i}{\kappa \pm k} \right) ^ {2+iZ(1/\kappa \pm1/k)}  \Gamma[2+iZ(1/\kappa \pm1/k)],
		\label{Jint}
\end{align}
where we have used an integral representation of a Gamma function $\Gamma(z)$ with complex argument.
In our case,  the contribution of the $J_+$ integral is vanishingly small as compared to that of  $J_-$. 
This is due to the IR photon energy being small compared to the final  kinetic energy of the electron, $\omega = k^2 /2 - \kappa^2/2 \ll k^2 /2$, so that the difference $|\kappa - k| \approx \omega/ k$ is much smaller than the sum  $\kappa + k \approx  2 k \pm \omega /k$. As a result, the fast oscillations of $\exp [i(\kappa  + k)r]$ lead to a relative cancellation of the corresponding integral, as compared to the one containing the factor $\exp [i(\kappa  - k)r]$. Neglecting the $J_+$ contribution, the asymptotic expression reduces to:
\begin{align}
T_{L,\lambda,\ell_i}(k;\epsilon_{\kappa})  \approx {}&
\frac{\pi}{2}~
N_{k} N_{\kappa}~
\braOket{R_{\kappa,\lambda}}{r}{R_{n_i,\ell_i}}  \nonumber \\ \times~&
\frac{ 1 }{ |\kappa-k|^2 } ~ \exp\left[-\frac{\pi Z}{2} \left( \frac{1}{\kappa}-\frac{1}{k}\right)\right]   
\nonumber \\
\times ~ & 
i^{L-\lambda-1}\exp[i(\eta_{\lambda}(\kappa)-\eta_{L}(k))]
\nonumber \\
\times ~ &
\frac{(2\kappa)^{iZ/\kappa}}{(2k)^{iZ/k}}
\frac{\Gamma[2+iZ(1/\kappa-1/k)]}{(\kappa-k)^{iZ(1/\kappa-1/k)}},
\label{Tas1}
\end{align}
which was used by us \cite{Klunder2011} to obtain estimates of the phases occurring in two-photon transitions entering RABBIT transition amplitudes. The first two lines in Eq.~(\ref{Tas1}) are real, they contain a one-photon matrix element from the bound state into the continuum, but also an exponential factor describing the strength of the continuum--continuum transition from 
$\kappa$ to $k$. 
The exponential factor decreases with the probe photon energy, $\omega = k^2/2-\kappa^2/2$, 
which indicates that large energy leaps in the continuum are strongly suppressed. 
At a given laser probe energy, however, the exponential factor increases with the final momentum, $k$,
which indicates that it becomes easier for the photoelectron to interact with the probe field.  
The third line is a simple phase factor containing the scattering phases of the continuum states.
Finally, the fourth line is a complex factor that depends on three quantities: final momentum, $k$; the laser probe frequency, $\omega$; and the charge of the ion, $Z$. 

A more formal derivation of this result, based on a closed-form representation of the Coulomb Green's function is given in the Appendix A. 
[We have found a typo in Eq.~(7) in ref.~\cite{Klunder2011}: The ratio, $(2k)^{i/k}/(2k_a)^{i/k_a}$, should be inverted, as is evident from Eq.~(\ref{Tas1}) in the present work].

Thus, in the asymptotic limit, the {\em phase} of the radial component takes the form: 
\begin{align}
\arg {[T_{L,\lambda,\ell_i}(k;\epsilon_\kappa)] } 
\approx {}&  
{\pi\over 2} (L-\lambda - 1)
\nonumber \\ 
+~& \eta_\lambda (\kappa) - \eta_L(k) 
+ \phi_{cc}(k,\kappa),	
\label{argTas1}
\end{align}
where 
\begin{align}
\phi_{cc}(k,\kappa)~ =~ \arg \left[  \frac{(2\kappa)^{iZ/\kappa}}{(2k)^{iZ/k}}~ 
{\Gamma[2+iZ(1/\kappa-1/k)] \over ( \kappa-k)^{iZ(1/\kappa-1/k)}}\right],
\label{phicc}
\end{align}
is the phase associated to a continuum--continuum radiative transition resulting from the absorption of $\omega$, in the presence of the Coulomb potential, $Z$.  It is important to note that it is {\it independent} from the characteristics of the initial atomic state, as well as from the amplitude of the field.
It is illustrative to study the continuum--continuum phase 
in the limit of a small photon energy, $\omega\approx k(k-\kappa)$, which yields a simplified expression: 
\begin{align}
\phi_{cc}^{(soft)}(k;\omega)~=~\arg\left[
\left( \frac{2k^2}{\omega} \right)^{iZ\omega/k^3} ~
\Gamma[2+iZ\omega/k^3]
\right],
\label{phiccsoft}
\end{align}
where it becomes clear that it is the product: $Z\omega/k^3$, which determines the size of the continuum--continuum phase.
This expression is expected to be valid in the so-called ``soft-photon'' limit, $k^2/2 \gg \omega $, 
where the exchange of  energy $\omega$ and the corresponding momentum transfer $\Delta k = \omega/c$  do not significantly modify the electron state \cite{Maquet}. The substitution, $\omega\rightarrow-\omega$, yields the (Fourier) component corresponding to stimulated emission of light. We note that the phases corresponding to absorption and emission have  opposite signs, but that they are otherwise identical in the soft-photon limit.

Replacing the formula obtained in Eq.~(\ref{Tas1}) for the radial component in the expression of the full transition amplitude $M ({\vec k},\epsilon_i+\Omega)$ given in Eq.~(\ref{Ma}), one gets its general form in the asymptotic limit:  
\begin{align}
M({\vec k};\epsilon_\kappa)
 \approx {}&  
-\frac{2\pi^2}{3}(8\pi)^{3/2}E_{\omega} E_{\Omega} N_k N_{\kappa}
\nonumber \\ \times~&
\frac{1}{|k-\kappa|^2}~\exp\left[-\frac{\pi Z}{2} \left( \frac{1}{\kappa}-\frac{1}{k}\right)\right]
\nonumber \\ \times~&
\frac{(2\kappa)^{iZ/\kappa}}{(2k)^{iZ/k}}
\frac{\Gamma[2+iZ(1/\kappa-1/k)]}{(\kappa-k)^{iZ(1/\kappa-1/k)}}
\nonumber \\ \times~&
\sum_{L=\ell_i,\ell_i\pm 2}  Y_{L,m_i}({\hat k})   
\sum_{\lambda=\ell_i\pm 1}\langle Y_{L,m_i}| Y_{1,0}|Y_{\lambda,m_i} \rangle  
\nonumber \\ \times~&
 \langle Y_{\lambda,m_i}| Y_{1,0}|Y_{\ell_i,m_i}\rangle
\braOket{R_{\kappa,\lambda}}{r}{R_{n_i,\ell_i}} i^{-\lambda} e^{i\eta_\lambda (\kappa)}  
\label{Mas1}
 \end{align}
To address the question of its phase, one notices that besides a trivial contribution from the spherical harmonic in the final state, 
$Y_{L,m_i}(\hat k)$, it contains only phase-shifts that are governed by the angular momentum $\lambda$ of the intermediate state, {\it i.e.} a state that can be reached via {\it single}-photon ionization. More precisely,  for a given transition channel characterized by the angular momenta of the intermediate and final state $\ell_i \rightarrow \lambda \rightarrow L$ , the phase of the matrix element reduces to: 
\begin{align}
\arg{ [M_{L,\lambda,\ell_i}({\vec k},\epsilon_\kappa)}] ={}&
\pi+\arg[Y_{L,m_i}({\hat k})] + \phi_\Omega + \phi_\omega 
\nonumber \\ -~&
{\pi \lambda \over 2} + \eta_\lambda (\kappa)  + \phi_{cc}(k,\kappa) ,	
\label{phiMas1}
\end{align}
 where $\phi_\Omega$ and $\phi_\omega$ are the phases of the XUV field $\Omega$ and of the IR laser $\omega$, respectively. 
We stress that the final state scattering phase, $\eta_L(k)$, cancels out and that it enters neither in  Eq.~(\ref{Mas1}) nor in Eq.~(\ref{phiMas1}). 
 
 Eq.~(\ref{phiMas1}) represents one of the major results of our theoretical analysis. It shows that, within the asymptotic approximation and besides trivial spherical harmonic contributions and the phases of the fields [line 1 in Eq.~(\ref{phiMas1})], the phase of a two-color ATI transition amplitude has two components: i) One is directly linked to the quantum-mechanical phase-shift of the one-photon XUV ionization amplitude, here $-\pi\lambda/2+\eta_\lambda (\kappa)$; ii) The other, denoted $\phi_{cc}(k,\kappa)$, is in some sense ``universal'', it describes the phase brought by the absorption of the probe photon $\omega$, in the presence of the Coulomb potential with charge $Z$.  Then, as shown below, when comparing laser-assisted ionization originating from distinct atomic states, the energy derivative of the phase-shifts in the first term contribute to a Wigner-like time-delay. On the other hand, the difference between the  ``universal'' terms $\phi_{cc}$ gives rise to a measurement-induced delay, associated to continuum--continuum stimulated radiative transitions in the presence of the Coulomb potential of the ionic core. 

\begin{figure}[h]
	\centering
		\includegraphics [width=0.45\textwidth]{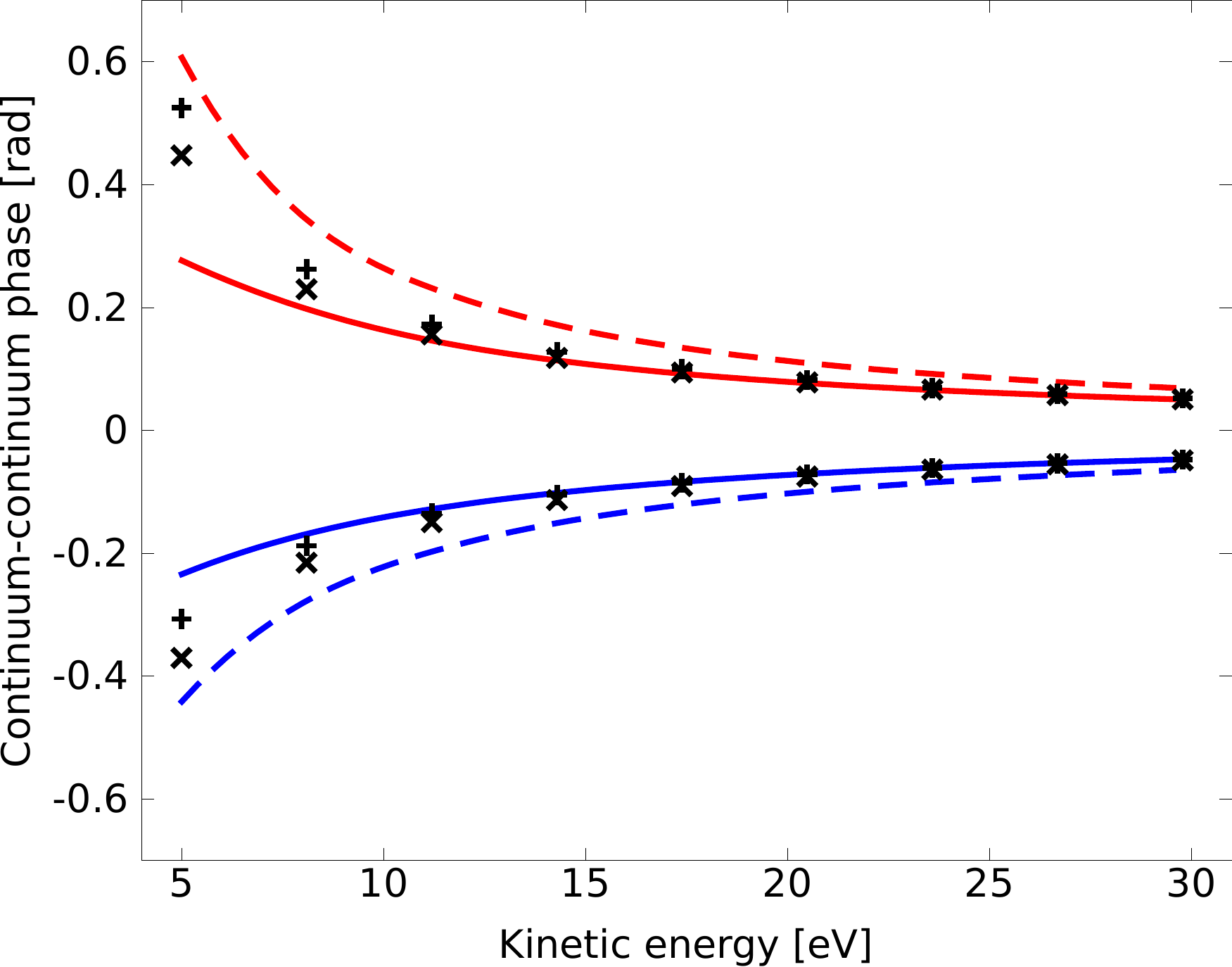}
	\caption{\label{results1a}
Continuum--continuum phases for absorption (red, upper curves) and emission (blue, lower curves)
calculated using the asymptotic approximation (dashed curves) [Eq.~\ref{phicc}] 
and the long-range amplitude-corrected asymptotic approximation (full curves) [Eq.~\ref{phicctilde}].
These approximate phases are compared with the exact calculations (black symbols) from the $1s$ state in hydrogen. 
The exact results for final angular momentum $L=0$ ($+$ symbol) and $L=2$ ($\times$ symbol)
are computed by subtracting the one-photon scattering phase from that of the exact two-photon matrix elements. 
The data correspond to $Z=1$ and to a laser probe with $\omega=1.55$~eV. 
It is interpolated between the discrete harmonic orders.
}
\end{figure}

In Fig.~\ref{results1a}, we present the continuum--continuum phases, 
associated with absorption (red) and emission (blue) of a probe photon
leading to the same final energy, calculated using 
the asymptotic approximation, Eq.~(\ref{phicc}). 
The continuum--continuum phases for probe photon absorption are positive while those for stimulated emission are negative, 
but approximately equal in absolute value.
In the next subsection, we show that, still in a single-active electron picture, it is feasible to improve the accuracy of our approximate treatment with the help of semi-classical arguments. 

\subsection{Long-range amplitude effects}

In order to go to the next level of our asymptotic approximation, 
we must include not only long-range phase variations of the continuum states,  
but also long-range variations of the amplitudes.   
Indeed, the normalization constants contained in the asymptotic forms of the radial functions  $R_{k,L}(r)$ and 
$\rho_{\kappa,\lambda}( r)$ can be modified to account for the long-range influence of the Coulomb potential. For instance, the modified final state normalization constant is: 
\begin{align}
N_k(r)=\sqrt{\frac{2}{\pi p(r)}},
\label{NWKB}
\end{align} 
where 
\begin{align}
p(r)=\sqrt{2(\epsilon-V(r))}\approx k - V(r)/k,
\label{pWKB}
\end{align}
is the local momentum from  Wentzel--Kramers--Brillouin (WKB) theory \cite{HaraldFriedrich}. The same remark applies to $N_{\kappa}(r)$ for the perturbed radial function.
Long-range amplitude effects can be approximated by expanding the quantities $N_k(r)$ and $N_{\kappa}(r)$, 
\begin{align}
N_k(r)N_{\kappa}(r) \approx 
\sqrt{\frac{4}{\pi^2 k \kappa}}
\left[
1-\frac{1}{2} 
\left(\frac{1}{\kappa^2} + \frac{1}{k^2} \right) \frac{Z}{r} 
\right], 
\label{normalizationOfr}
\end{align}
to first-order in the Coulomb potential. 
The second term within the brackets in Eq.~(\ref{normalizationOfr}) contains the first-order amplitude correction to the matrix element. Evaluation of the long-range amplitude contribution leads to a correction to the continuum--continuum phase:
\begin{align}
\alpha_{cc}(k,\kappa) ={}&
\arg\left[ 1 + 
\frac{iZ}{2} 
\left( \frac{1}{\kappa^2} + \frac{1}{k^2} \right )
\frac { \kappa - k }{1+iZ(1/\kappa-1/k)}\right].
\label{ampCorrFactor}
\end{align}
The final continuum--continuum transition phase is:
\begin{align}
\tilde {\phi}_{cc}(k,\kappa) = \alpha_{cc}(k,\kappa) + \phi_{cc}(k,\kappa) 
\label{phicctilde}
\end{align}
where $\phi_{cc}(k,\kappa)$ is given in Eq.~(\ref{phicc}).
In Fig.~\ref{results1a}, we show that including such long-range effects improves the accuracy of the approximation greatly, leading to accurate  quantitative result already at relatively low energies in the continuum. The accuracy in the lower energy range  can be further improved by using a regularization method designed to remove the effect of the singularity in Eq.~(\ref{normalizationOfr}) as $r$, $k$ and $\kappa$ go to zero. 

Going beyond the approximations given here, namely performing {\it exact  ab initio} computations of the matrix elements $M({\vec k}, \epsilon_i+\Omega)$ in polyelectronic systems, is out of reach of present computational capabilities.  It is only in the special case of hydrogenic systems, that such 2-photon amplitudes can be computed with arbitrary precision. 
Thus, with objective to  delineate the range of validity of our approximation, 
we turn now to a brief presentation of the ``exact'' calculations in hydrogen.

\subsection{Exact calculations of 2-photon ATI matrix elements in hydrogenic systems}

The principle of the calculation is outlined here for  $s$-states. Numerical data for other states will be given below. We first  express the transition amplitude given in Eq.~(\ref{Ma}) for the case $ \ell_i = 0, m_i = 0$ which implies $\lambda = 1$ so that the angular momentum of the photoelectron is either $L= 0,2$. Accordingly, two distinct amplitudes contribute to  ATI transitions like the one depicted in   Fig.~\ref{fig1}~(c):
\begin{align}
\label{Mas}   
\left. M({\vec k},\epsilon_\kappa)\right|_{\ell_i=0} ={}&
\:{1 \over 3i} (8 \pi)^{3/2}  E_\omega E_\Omega 
\nonumber \\ \times~&
\left[ e^{ i \sigma_{0}(k)} Y_{0,0}({\hat k})   T_{0,1,0}(k; \epsilon_\kappa) \right.
\nonumber \\ -~&
\left. {2 \over \sqrt{5} }e^{ i \sigma_{2}(k)} Y_{2,0}({\hat k})  T_{2,1,0} (k; \epsilon_\kappa)\right] ,
\end{align} 
 where the radial components for $s-$states are of the form:
 \begin{align}
\left. T_{L,1,0} (k;\epsilon_\kappa)\right |_{\ell_i=0}={}& 
\langle R_{k,L}| r |G_1 (\epsilon_\kappa)| r |R_{n_i,0} \rangle	,
\label{TGreen}
 \end{align}
 with $L = 0,2$; $n_i$ labels the initial atomic $s-$state and $G_{\lambda = 1} (\epsilon_\kappa)$ is the radial component of the Coulomb Green's function for angular momentum $\lambda =1$. The general form of the Green's function with energy argument $\epsilon_\kappa$ is:
\begin{align}
 \label{Green}  
G_\lambda (r',r;\epsilon_\kappa) =  
\lim_{ \varepsilon \to  0^+} \int \!\!\!\!\!\!\!\! \sum_{\nu}  
{R_{\nu,\lambda}(r') R_{\nu,\lambda} (r)  \over \epsilon_\kappa -\epsilon_\nu  +i \varepsilon} .  
\end{align} 
As already mentioned, the infinite sum over the index $\nu$ runs over the whole (discrete + continuous) spectrum of the hydrogenic system. Closed form expressions for $G_\lambda$ are known, see for instance \cite{Maquet98}. Here, we have used the expression given as an expansion over a discrete Sturmian basis:
\begin{align}
     \label{GreenSturm}  
G_\lambda (r',r;\epsilon_\kappa)=   \sum_{\nu=\lambda +1}  {S_{\nu,\lambda,x}(r') S_{\nu,\lambda,x} (r)  \over 1 - \nu x} , 
\end{align} 
 where $x = \sqrt{-2\epsilon_\kappa}$ and the so-called Sturmian functions $S_{\nu,\lambda,x} (r)$ have a structure similar to the  bound-state hydrogenic radial functions \cite{Rotenberg,Maquet1977}:  
\begin{align}
     \label{Sturm}  
S_{\nu,\lambda,x}(r)  ={}& 2x\sqrt{(\nu -\lambda -1)! \over(\nu +\lambda)!} 
\nonumber \\ \times ~ &
e^{-xr} (2xr)^\lambda L_{\nu - \lambda-1}^{2\lambda +1} (2xr),
\end{align} 
where $L_{\nu - \lambda-1}^{2\lambda +1} (z)$ are associated Laguerre polynomials. 
In the amplitudes for ATI transitions, $\epsilon_\kappa =  \epsilon_i +\Omega >0$, and the quantity, $x =i \sqrt{2|\epsilon_\kappa|}$, is pure imaginary. It is then convenient to use Pad\'e-like resummation techniques to compute the infinite sum over the index $\nu$, see, for instance ref. \cite{Klarsfeld79b}.  

In Fig.~\ref{results1a}, we present the exact continuum--continuum phases from the $1s$ state in hydrogen.
These phases are defined as the total phase of the exact matrix element, 
$M_{L,1,0}(\vec k;\epsilon_\kappa)$,  
{\it minus} the one-photon phase [see line~2 of Eq.~(\ref{phiMas1})]. 
Our approximate calculation including long-range amplitude effects, Eq.~(\ref{phicctilde}),
is in excellent agreement with the exact calculations except at low energy.
%

\subsection{Phase of the classical dipole}
\label{sec:classical}

Finally, we present a simplified derivation of the continuum--continuum phase, 
$\phi_{cc}(k,\kappa)$, using a classical approach.
The dipole associated with the absorption of radiation at frequency $\omega$ by a free electron in the presence of a Coulomb potential, can be calculated using Larmor's formula, 
\begin{align}
d_C(k;\omega)=\int_0^{\infty} dt ~ r_k(t) \exp[- i\omega t],
\label{cl_dipole}
\end{align}
where it is assumed that the electron follows a {\it field-free} trajectory, $r_k(t)$, that starts close to the ion, $r_k(0) \approx 0$, and then moves out away from the ion with an asymptotic velocity, $k$. 
The integral can be cast from time to space using the $r-$dependence of the velocity:  
\begin{align}
v_k(r)~ =~ \sqrt{k^2  - 2 V(r)}, 
\label{cl_time2space}
\end{align}
where $k^2/2$ is the final kinetic energy of the electron at large distance from the ion. 
Using the differential $dt = dr/v(r)$, the time can be written as 
$t(k;r) = \int^r dr'/v_k (r')+C$, where $C$ is an integration constant. 
In the case of the Coulomb potential, $V(r)=-Z/r$, the integral becomes
\begin{align}
t(k;r) ={}&
\int^{r}dr' ~ \frac{1}{ \sqrt{k^2+2 Z/r'} }  + C 
\nonumber \\ \approx{}&
 \frac{r}{k}-\frac{Z}{k^3}\ln(r)+C,
\label{CoulombTimeApprox}
\end{align}
in the asymptotic limit, {\it i.e.} when $k^2 /2 \gg Z/r$. This provides an approximate time--position relation valid at large distances from the origin. 
In the special case where the electron starts from the origin $[t,r]=[0,0]$, 
the exact integration in Eq.~(\ref{CoulombTimeApprox}) leads to  $C=-Z\ln[2k^2/Z]/k^3$. Keeping for the moment this value of the constant and replacing the asymptotic form of the time in the expression of the dipole Eq.~(\ref{cl_dipole}), one gets:
\begin{align}
d_C(k;\omega) \approx{}& 
\frac{1}{k}\int_0^{\infty}dr ~ r\exp\left[ - i\frac{\omega}{k}\left(r-\frac{Z}{k^2}\log[2k^2r/Z]
\right) \right] 
\nonumber \\ {}=& 
\frac{1}{k}\left(\frac{2k^2}{Z}\right)^{ iZ\omega/k^3}  \int_0^{\infty}dr ~ 
r^{1 + iZ\omega/k^3}\exp\left[- i\frac{\omega}{k} r\right] 
\nonumber \\ ={}&
-\frac{k}{\omega^2}\exp\left[-\frac{3 \pi Z \omega}{ 2k^3}\right]
\nonumber \\ \times~& 
\left( \frac{2k^3}{Z\omega} \right)^{iZ\omega/k^3}
\Gamma(2+iZ\omega/k^3),
\label{C_dipole}
\end{align}
where the next-to-last line is real, with an exponential factor that decreases with $\omega$, but increases with $k$,
in excellent agreement with the quantum mechanical result, Eq.~(\ref{Mas1}).
Furthermore, the last line contains the complex gamma function times an algebraic factor,
also in close connection to the quantum counterpart.  Clearly, the dipole corresponding to absorption is a complex quantity with phase: 
\begin{align}
\phi_C(k;\omega) ={}&
\arg\left[d_C(k;\omega)\right]  
\nonumber \\ \approx{}&
\arg\left[-\left( \frac{2k^3}{Z\omega} \right)^{iZ\omega/k^3}~\Gamma(2+iZ\omega/k^3) \right],
\label{C_phase}
\end{align}
which is closely related, but {\it not} identical to its soft-photon quantum counter part $\phi^{(soft)}_{cc}$,   
given in Eq.~(\ref{phiccsoft}).

In the quantum mechanical case, the electron starts in a bound state with some spatial extent and not exactly from $r=0$. In order to account for this uncertainty on the initial position, we may choose a different value of the integration constant $C$, introduced in Eq.~(\ref{CoulombTimeApprox}), in order to come closer to the quantum mechanical dipole. This matching-procedure is reminiscent of the method used in ref.~\cite{IvanovM2011}, to determine the ``best'' initial radial position for the electron within the eikonal Volkov approximation. Within the lowest-order approximation of Eq.~(\ref{CoulombTimeApprox}), we find a simple relation between the initial position and the integration constant: 
\begin{align}
r_0 \approx \exp\left[\frac{Ck^3}{Z}\right], 
\label{r2C}
\end{align}
which is valid at high-energy.
It is convenient to set $C=-Z\ln[2k]/k^3$, 
a choice corresponding to an initial radial position $r_0\approx 1/(2k)$, as was identified in ref.~\cite{IvanovM2011}.
Clearly, when the first-order amplitude correction is included,  Eq.~(\ref{phicctilde}), 
the initial position is adjusted accordingly. 
The continuum--continuum phase being only one part of the total quantum mechanical phase, 
we have also to include the scattering phase if we want to deduce the ``true'' initial position. 
Interestingly, in our approach this exact inital position is not critical.
In fact, our results are stable with respect to rather substantial modifications of the wavefunctions close to the core.
It is the behavior of the wavefunctions far away from the core that must be described accurately, using the asymptotic approximation. 
We now turn our attention to the applications of the complex ATI matrix elements, to the determination of attosecond delays in photoionization.

\section{Attosecond Time-delays}
\label{sec:ATIapplications}
The complex amplitude $M({\vec k},\epsilon_i+\Omega)$ 
(denoted $M^{(a)}$ for conciseness in the following) 
for the joint absorption of an XUV photon $\Omega$ and of an IR  laser photon $\omega$,
is of course {\it not} a direct observable in any experiment. 
Only the square modulus of a complex transition amplitude can be measured. 
Thus, if $M^{(a)}$ is the only amplitude leading to a given final state,  there is no way to determine its absolute phase. However, in the cases of interest here, with attosecond XUV pulses in the presence of a probe IR laser field, several other channels are open which can lead to the same final state, thus making it feasible to observe phase-dependent interference patterns. This property is exploited in the following schemes.

\subsection{Attosecond delay measurements using pulse trains}

In the RABBIT scheme there are two dominant complex amplitudes (quantum paths), $M^{(a)}$ and $M^{(e)}$, 
associated with the absorption of harmonic $H_{2q-1}$ or $H_{2q+1}$ plus absorption or emission of a laser photon
with phase, $\pm\omega\tau\equiv\pm\phi_\omega$, 
leading to the same final sideband, $S_{2q}$. 
The photoelectron will transit via different intermediate states: $\kappa_<$ and $\kappa_>$,
corresponding to different intermediate energies:
$\epsilon_<=\epsilon_i+2q\omega-\omega=\kappa_<^2/2$ 
and
$\epsilon_>=\epsilon_i+2q\omega+\omega=\kappa_>^2/2$. 
Clearly, the energy of the corresponding intermediate states are one photon below and above the sideband, as is illustrated in Fig.~\ref{fig1}~(a). 
This implies that the measured intensity of the sideband, $S_{2q}$, will depend on 
the {\it phase difference} between the two quantum paths:
\begin{align}
P_{2q} \propto | M^{(a)}+M^{(e)} |^2  =  |M^{(a)}|^2 + |M^{(e)}|^2 \nonumber \\
 + ~ 2 |M^{(a)} | | M^{(e)}| \ \cos\left[\arg\left(M^{(a)*} M^{(e)}\right)\right] 
\label{SBprob}
\end{align} 
which corresponds to a standard interference phenomenon, as summarized in Eqs. (\ref{S2q}) or, equivalently in  (\ref{S2qt}) displayed in the introduction (Sec.~\ref{sec:intro}).
If we now apply Eq.~(\ref{phiMas1}), and assume that the total contribution to the complex amplitudes  
can be approximated by a single intermediate angular momentum component, $\lambda$, 
the phase difference is 
\begin{align}
\arg\left(M^{(a)*}_{L,\lambda,\ell_i}  M^{(e)}_{L,\lambda,\ell_i}\right)  \ \approx \ 
-2\omega\tau ~+~ 
\overbrace{
\phi_{2q+1}-\phi_{2q-1} 
}^{\Delta\phi_{2q}}
\nonumber \\ 
~+~
\underbrace{
 \eta_{\lambda}(\kappa_>) 	-\eta_{\lambda}(\kappa_< )
~+~ \phi_{cc}(k,\kappa_>)  		-\phi_{cc}(k,\kappa_<)
}_{\Delta\theta_{2q}}.
\label{phaseDiff}
\end{align}
Here the first line contains the phases of the fields, 
while the second line is $\Delta\theta_{2q}$, 
in terms of scattering and continuum--continuum phases. 
The corresponding measurement delay, $\tau_\theta$ as defined in Eq.~(\ref{S2qt}),  
is the sum of finite-difference approximations to a Wigner-like time-delay: 
\begin{align}
{\tau}_\lambda(k) \approx \frac{ \eta_\lambda (\kappa_>)- \eta_\lambda (\kappa_< )}{2\omega} 
\label{Wigner}
\end{align}
and to a continuum-continuum delay:
\begin{align}
\tau_{cc}(k;\omega) \approx \frac{\phi_{cc}(k,\kappa_>) - \phi_{cc}(k,\kappa_<)}{2\omega}, 
\label{Taucc}
\end{align}
so that $\tau_\theta \approx { \tau}_{\lambda} + \tau_{cc}$. 

In a more general case, 
the interference will result from complex amplitudes with two terms for absorption and for emission,
$M^{(a/e)}=M^{(a/e,+)}+M^{(a/e,-)}$, 
corresponding to different intermediate angular momenta, $\lambda=\ell_i \pm 1$. The interesting point is that
the $\tau_{cc}$ will not be affected, since the phases $\phi_{cc}$ are independent of $\lambda$ as well as of the final angular momenta $L=\ell_i, \ell_i \pm 2$, see Eq.~(\ref{phicc}).
The remaining part can be interpreted 
as an {\it effective} Wigner-like delay, $\tilde {\tau}_{\lambda = \ell_i \pm 1}$, which has to be computed taking into account the relative weights of the {\it four} components entering the expression of the transition probability amplitude.

\subsection{Attosecond delay measurements using single pulses }

\begin{figure}[h]
	\centering
		\includegraphics [width=0.35\textwidth]{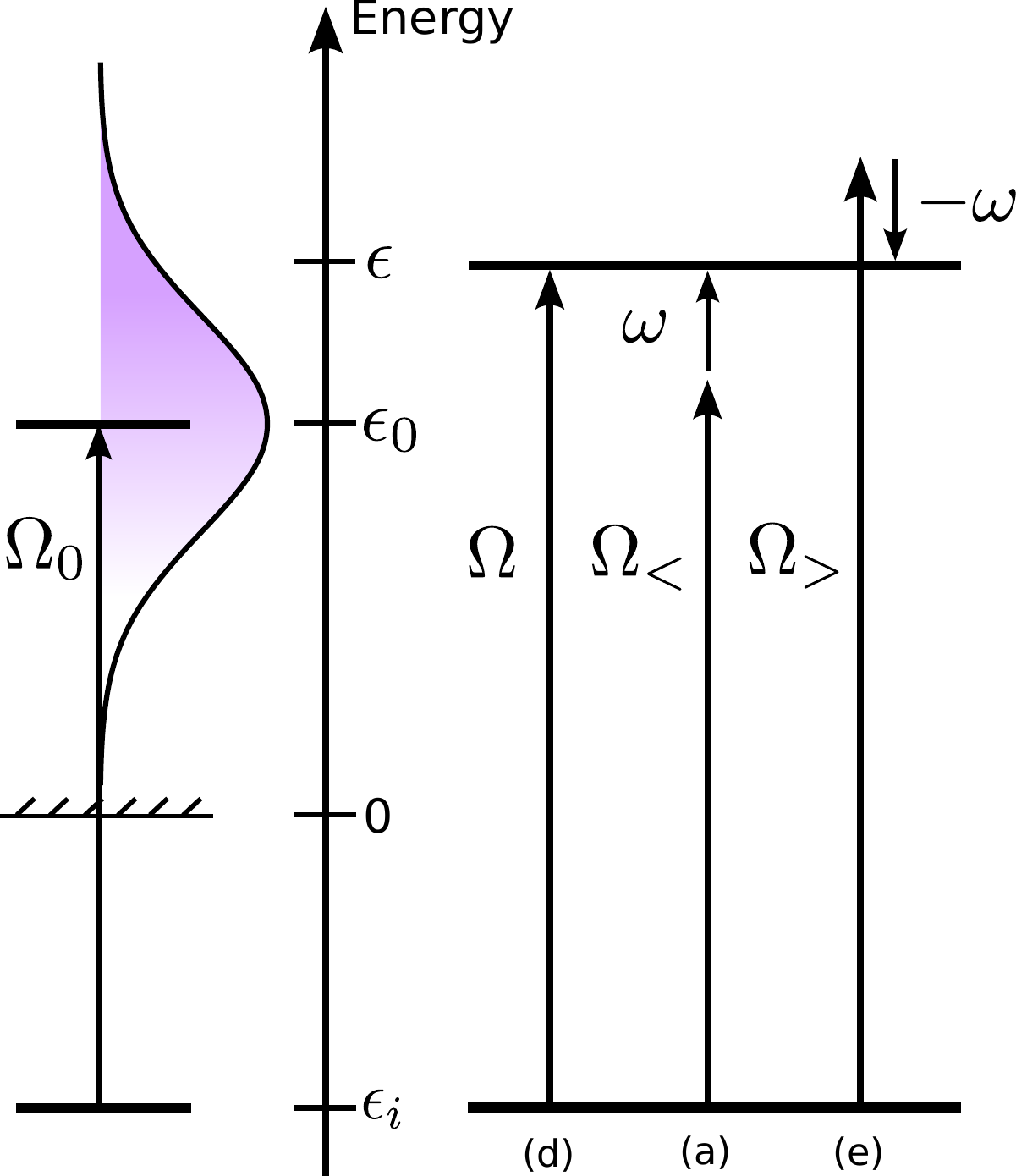}
	\caption{\label{fig4}
Sketch of the quantum paths describing the onset of ``streaking'' for a photoelectron 
ionized by a single attosecond pulse and probed by a monochromatic laser field.
The first-order photoelectron wave packet is centered at $\epsilon_0=\epsilon_i+\Omega_0$.
Within the bandwidth of the attosecond pulse, any energy, $\epsilon$, can be reached by path~(d), 
where a single XUV photon with frequency $\Omega$ is absorbed. 
Alternatively, 
the same energy can reached by path~(a) 
by absorbing a less energetic photon, $\Omega_{<}$, and a laser photon, $\omega$;
or by path~(e) 
by absorbing a more energetic photon, $\Omega_{>}$, and then emitting a laser photon, $-\omega$. 
At the high-energy end of the photoelectron distribution (indicated by $\epsilon$), 
the dominant contributions to the photoelectron wave packet are (d) and (a), 
due to $\Omega_{>}$ being far in the upper energy range of the XUV bandwidth.   
	}
\end{figure}

In this section we apply the perturbative treatement to laser-assisted photoionization 
by a single-attosecond pulse (SAP) and a probing laser field. 
The ionizing attosecond field is 
\begin{align}
\tilde E_{SAP}(t)=\int d\Omega ~ E_\Omega \exp[-i\Omega t]/2\pi,
\label{SAP}
\end{align} 
with $E_\Omega=|E_\Omega|~\exp[i\phi_\Omega]$ being complex Fourier coefficients. 
While our approach is very general, 
it can be illustrated by considering a Gaussian frequency distribution centered on the frequency $\Omega_0$, 
as depicted in the left panel of Fig.~\ref{fig4}. 
The probing laser field is assumed to be monochromatic and real, $\tilde E(t)=2|E_\omega|\cos[\omega (t - \tau)]$,
so that it can account for both absorption and emission processes. 
Our perturbative approach cannot be used to fully account for the   
large momentum shifts that are typical of streaking spectrograms, but
it does allow us to study the \textit{onset} of streaking. 
As we shall see below, streaking-like behavior 
is clear already in the perturbative regime. 
In the right panel of Fig.~\ref{fig4}, we sketch the interfering processes to a certain energy $\epsilon$: (d), (a) and (e) from lowest-order perturbation theory. The dominant contribution (d) corresponds to absorption of a single XUV photon, while the upshifted (a) and downshifted (e) photoelectron spectra corresponds to absorption and emission of an additional laser photon $\omega$, respectively.


The matrix element for {\it one-photon} ionization with a photoelectron emitted along the polarization axis is
\begin{align}
M^{(d)}(k\hat z;\epsilon_i+\Omega) ={}&
\frac{(8\pi)^{3/2}}{i}
E_\Omega 
\sum_{\lambda}
\left(-i\right)^{\lambda}  
\sqrt{\frac{2\lambda+1}{3}}
e^{i\eta_{\lambda}(k)}
\nonumber \\ \times~&
\braOket{Y_{\lambda,0}}{Y_{1,0}}{Y_{\ell_i,0}}
\braOket{R_{k,\lambda}}{r}{R_{n_i,\ell_i}}~
\delta_{m_i,0},
\label{Lforward}
\end{align}
where the XUV frequency satisfies $\epsilon_i+\Omega = \epsilon=k^2/2$ for process (d). Note that only initial states with zero magnetic quantum number,  $m_i=0$, will contribute to photoelectron emission along $\hat z$, as indicated by the Kronecker delta at the end of Eq.~(\ref{Lforward}) that originates from the explicit properties of 
the spherical harmonics along $\hat z$: $Y_{\lambda,m_i}(\hat z)=\sqrt{(2\lambda+1)/3}~\delta_{m_i,0}$.

The {\it two-photon} matrix element for photoelectrons along the polarization direction is 
\begin{align}
M(k\hat z;\epsilon_i+\Omega_{<})={}&
\frac{(8\pi)^{3/2}}{i}
E_{\Omega_<}E_\omega
\sum_{L}
\left(-i\right)^{L}
\sqrt{\frac{2L+1}{3}}
e^{i\eta_L(k)}
\nonumber \\ \times ~ &
\sum_{\lambda} 
\braOket{Y_{L,0}}{Y_{1,0}}{Y_{\lambda,0}} 
\braOket{Y_{\lambda,0}}{Y_{1,0}}{Y_{\ell_i,0}}  
\nonumber \\ \times ~ &
T_{L,\lambda,\ell_i}(k;\epsilon_i+\Omega_{<}) ~
\delta_{m_i,0}
,
\label{Mforward}
\end{align}
where $\epsilon_i+\Omega_{<}+\omega=\epsilon$ for process (a). 
A similar expression, $M(k\hat z,\epsilon_i+\Omega_{>})$, can be written for process (e) 
where $\epsilon_i+\Omega_{>}-\omega=\epsilon$, and where the complex conjugate of the laser field, $E_\omega^*$, 
is used for emission of a laser photon. Again, only initial states with zero magnetic quantum number contribute.

The PROOF method \cite{Chini2010} relies on the interference between these three quantum paths, denoted $M^{(d)}$, $M^{(a)}$ and $M^{(e)}$, as a function of the XUV-laser delay, $\tau$. 
The probability for emission of an electron with final momentum $k$ along the polarization axis $\hat z$ is 
\begin{align}
P(k \hat z) \propto |M^{(d)}+M^{(a)}+M^{(e)}|^2,
\label{streakingProb}
\end{align} 
where only cross-terms lead to delay dependent modulations.
Provided that the laser field is {\it relatively} weak, we expect the interference cross-terms (d)-(a) and (d)-(e), will dominate over the cross-term (a)-(e), because the former involve the exchange of only one laser photon while the latter contributes to higher order terms in a perturbative treatment,  involving the exchange of two laser photons $\omega$.
	
In order to illustrate the origin of the attosecond delay within this framework, we  consider a final energy that is higher than the central energy of the photoelectron wave packet, $\epsilon=\epsilon_i+\Omega > \epsilon_i+\Omega_0$.
In this region, the $\omega$ modulation is dominated by the (d)-(a) interference, 
because path (a) goes through a central part of the XUV distribution, 
while path (e) passes through a higher frequency range, in the upper part of the XUV bandwidth [Fig.~\ref{fig4}].
Increasing the bandwidth of the attosecond pulse or decreasing the probe photon frequency makes this distinction less pronounced.
We restrict our analysis to the case of an initial $s$-state, $\ell_i=0$, 
and study the interference of the cross term (d)-(a). 
The relevant phase reads
\begin{align}
\arg[M^{(a)*} M^{(d)}] \approx {} &
-\omega\tau +
\phi_\Omega-\phi_{\Omega_<}
\nonumber \\ +~&
\eta_{\lambda}(k)-\eta_{\lambda}(\kappa_{<})
-\phi_{cc}(k,\kappa_<)
-\frac{\pi}{2},
\label{LMcross term}
\end{align}
where we have used the asymptotic approximation for the phase of the two-photon matrix element as in Eq.~(\ref{phiMas1}), thereby,  introducing the continuum--continuum phase into the framework of laser-assisted photoioization by single attosecond pulses. 
Also, we note that the phase-shifts present in the two-photon matrix element  are those of $p-$waves ($\lambda=1$). In terms of the temporal delays, the $\omega$ modulation is displaced by
\begin{align}
\tau \approx {} & 
\frac{\phi_\Omega-\phi_{\Omega_<}}{\omega}
+\frac{\eta_{\lambda}(k)-\eta_{\lambda}(\kappa_<)}{\omega}
-\frac{\phi_{cc}(k,\kappa_<)}{\omega} - \frac{\pi}{2\omega} 
\nonumber \\ \equiv{}& 
\tau_\Omega ~+~ \tau_\lambda(k) ~+~ \tau_{cc}(k;\omega) ~-~ \frac{\pi}{2\omega},
\label{streakingdisplaced}
\end{align}	
where we have used the definition of the continuum--continuum delays $\tau_{cc}$, Eq.~(\ref{Taucc}), and the following relation for the continuum--continuum phases $\phi_{cc}$:
\begin{align}
\phi_{cc}(k,\kappa_<)\approx -\phi_{cc}(k,\kappa_>)
\label{ccrelation},
\end{align}
which is exact in the soft-photon limit.
Similarly to Eq.~(\ref{streakingdisplaced}), we can compute the modulation at the low energy end of the electron distribution  using the cross term (d)-(e), and we find the same result  but shifted by a half laser period, {\it i.e.} out of phase by $\pi$.  
This $\pi-$shift of the modulation is important to explain the onset of streaking, because it ensures that the high-energy probability of the electron spectra is maximized when the low-energy part is minimized. In the central region, $\epsilon\approx\epsilon_i+\Omega_0$, 
the (d)-(a) and (d)-(e) contributions will be comparable, leading to a relative cancellation of the $\omega$ modulation, leaving only the $2\omega$ modulation. In total, the electron momentum distribution is slightly shifted up or down depending on the sub-cycle delay between the laser field and the attosecond pulse: It is {\it streaked} as expected from the classical (or strong-field) picture. 
Using Eq.~(\ref{streakingdisplaced}), we find that this streaking modulation is displaced by a  Wigner-like delay, $\tau_{\lambda}(k)$, and by the continuum--continuum delay, $\tau_{cc}(k;\omega)$. We can interpret this delay as the time it takes for the electron to be photoionized {\it plus} the time it takes for the measurement process to occur, {\it i.e.} for a lower energy electron to absorb one probe photon so that it may interfere with the direct path. 
We stress that this analysis was made assuming a monochromatic probe field, 
while it is common in streaking experiments to use few-cycle IR laser pulse, 
which leads to a convolution (a blurring effect) of the momentum modulation and of the attosecond time-delays.

\subsection{Comparison between the two measurements}

We now briefly discuss the small differences of $\tau_{cc}$
occuring between the RABBIT method and the streaking method.  
Eq.~(\ref{ccrelation}) implies that 
\begin{align}
\tau_{cc}(k;\omega)  \equiv {} & 
\frac{\phi_{cc}(k,\kappa_>)~-~\phi_{cc}(k,\kappa_<)}{2\omega}
\nonumber \\  \approx {} &
~\frac{\phi_{cc}(k,\kappa_>)}{\omega},
\label{ccphases}
\end{align}
where the first line corresponds to the RABBIT method and 
the second line corresponds to the streaking method. 
These different ``flavours'' of $\tau_{cc}$ merge completely in the soft-photon limit, $\epsilon\gg \omega$,
but differ slightly at low kinetic energies,
where the phases are not exact opposites. 
In both methods, $\tau_{cc}$ varies with $k$ inside the bandwidth of the photoelectron wavepacket.

Furthermore, it is interesting to note that the group delays of the attosecond pulses and Wigner delays of the photoelectrons appear as discrete derivatives over one and two $\omega$ photons in the streaking and RABITT methods respectively. Here, we have assumed that all delays are slowly varying so that the discrete derivatives are equivalent.
We also note that the streaking method relies on the cross-terms: (d)-(a) and (d)-(e); 
while the RABBIT method relies on the (a)-(e) contribution. 
The weaker signal in the RABBIT method is not a problem, since it is recorded between the harmonics, on ``zero background''. Consequenty, the modulation frequency of the streaking/PROOF signal is $\omega$ due to the single laser photon involved in each appropriate cross term, while the corresponding RABBIT frequency is $2\omega$ due to the two laser photons involved in the latter cross term. 
Another important difference between the methods is related to symmetry (parity): The $\omega$ signal is somewhat restricted to the $\hat z$ direction and it is not observed if photoelectrons are collected in all directions; while the $2\omega$ signal is more general and it prevails also when electrons in all directions are collected. 
Interestingly, we have shown that all three attosecond characterization methods (RABITT, PROOF and streaking) 
provide equivalent temporal information about the XUV ionization process, 
even though they are built from different sets of cross-terms associated to different interfering quantum paths.

\section{Results}
\label{sec:results}

\subsection{Calculations of attosecond time-delays}

\begin{figure}[h]
    \centering
        \includegraphics [width=0.45\textwidth]{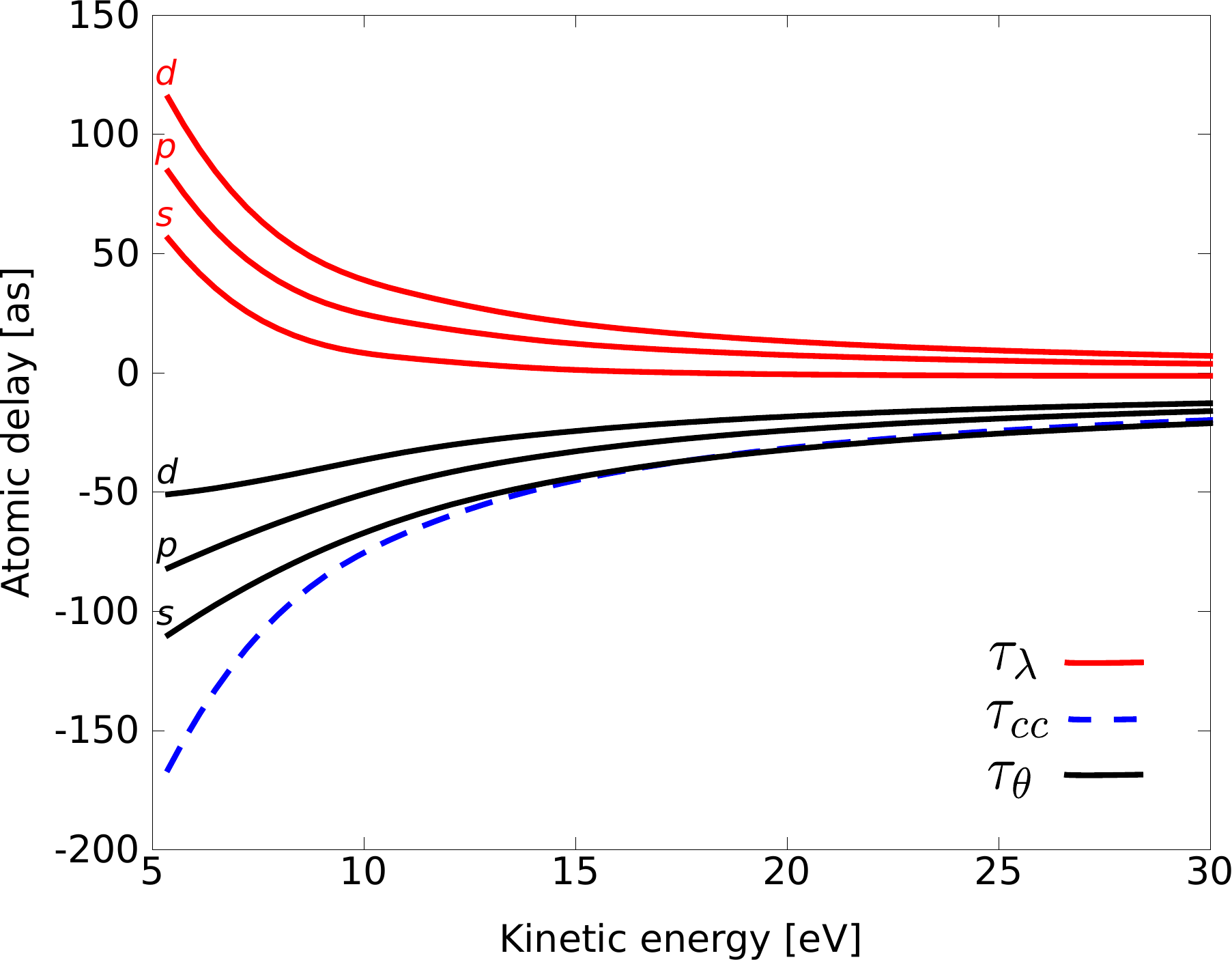}
    \caption{\label{results5}
Atomic delays of laser-assisted photoionization in hydrogen 
for three different angular momenta of the wave-packet wavefunction in the intermediate state ($s$, $p$ and $d$ : $\lambda=0$, 1 and 2). 
The delays are calculated using the regularized approximation.
The data correspond to $Z=1$ and a laser probe with $\omega=1.55$~eV.
}
\end{figure}

In Fig.~\ref{results5}, we plot atomic delays relevant for laser-assisted photoionization.
The Wigner-like delays, $\tau_{\lambda}$, are calculated from the finite-difference derivative of the scattering phase of hydrogen for
angular momentum, $\lambda=[0,1,2]$, corresponding to $s$, $p$ and $d$ continuum wave packets.
The universal continuum--continuum delay, $\tau_{cc}$, plus these Wigner-like delays yields the total atomic delay, $\tau_\theta$.
Notice that the sign of $\tau_\lambda$ and $\tau_{cc}$ are opposite,
so that $\tau_\theta$ is smaller than either of the contributions individually.
The increase of $\tau_\lambda$ with the angular momentum can be understood as due to the repulsive, short-range, centrifugal potential.
The total delay is negative in this example, which implies that the electron appears as being {\it advanced} compared to the probe field.

\begin{figure}[h]
	\centering
		\includegraphics [width=0.45\textwidth]{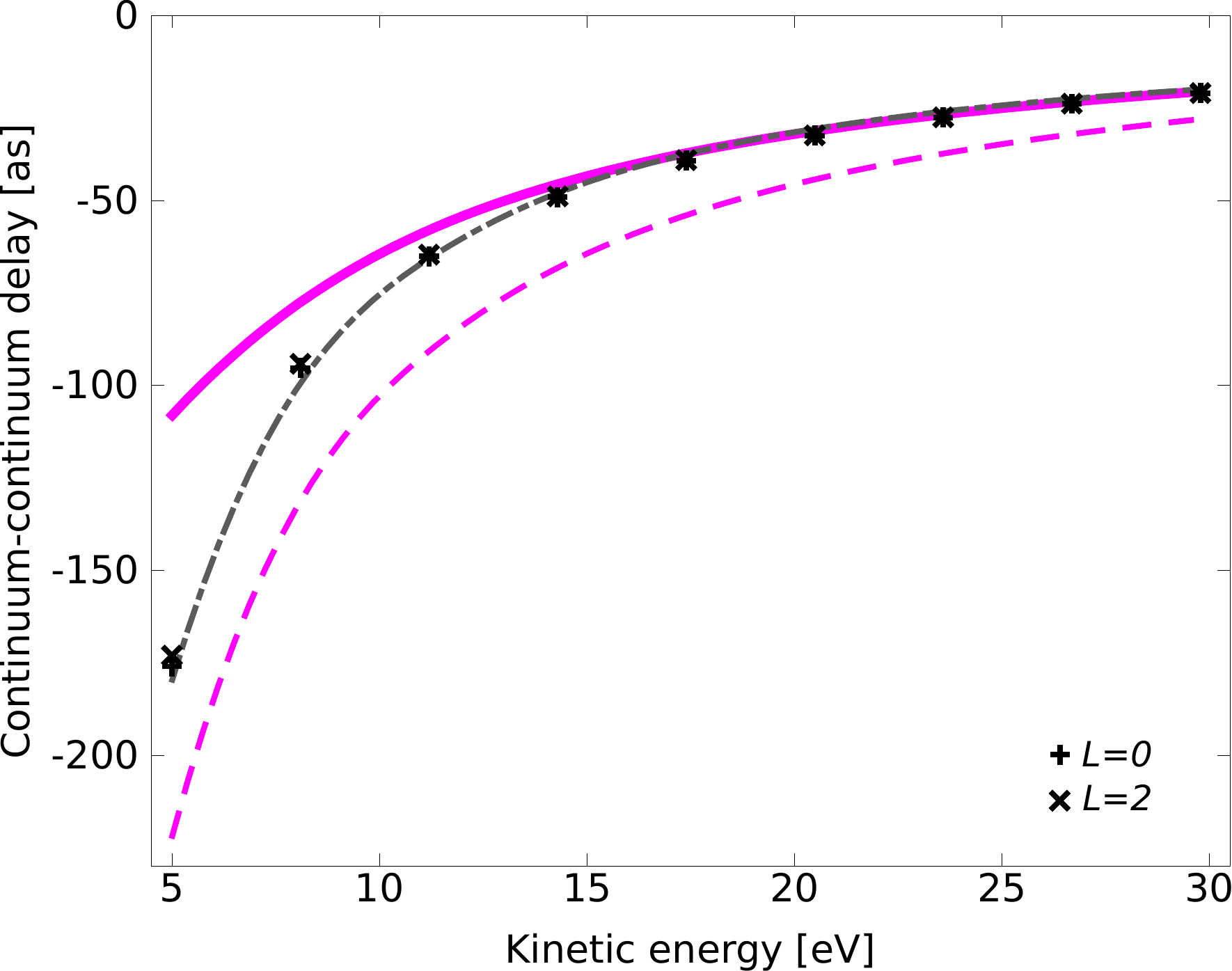}
	\caption{\label{results1b}
Continuum--continuum delays calculated using the asymptotic approximation (dashed curve),  Eq.~(\ref{phicc}), 
the long-range amplitude-corrected asymptotic approximation (thick curve), Eq.~(\ref{phicctilde}),
and the regularized asymptotic approximation (dash-dot curve).
The approximate delays are compared with exact delays computed from the $1s$ state in hydrogen 
by subtracting the intermediate Wigner delay. 
The data correspond to $Z=1$ and to a laser probe with $\omega=1.55$~eV.
}
\end{figure}

In Fig.~\ref{results1b}, we evaluate the accuracy of $\tau_{cc}$ 
as derived by various degrees of the asymptotic approximation by 
comparison with exact calculations in hydrogen from the $1s$ state along 
two different angular momentum sequences: 
$s \rightarrow p \rightarrow s$ ($+$) and $s \rightarrow p \rightarrow d$ ($\times$). 
%
Indeed, we find that the exact $\tau_{cc}$  
are almost completely independent of the final state angular momentum. 
The asymptotic approximation (dashed curve) predicts the correct qualitative behavior of the delay, 
but it slightly overestimates its magnitude. 
By taking into account the long-range amplitude effects, the agreement is excellent at high kinetic energies.
The disagreement at low energy can be removed by avoiding the radial singularity in Eq.~(\ref{normalizationOfr}), 
namely by an {\it ad hoc} substitution of $r \rightarrow r+i(1-|k-\kappa|/2)$. In this way, we obtain a ``regularized'' continuum--continuum delay 
(dot--dashed curve), which is excellent at {\it all} energies in the range.

\begin{figure}[h]
	\centering
		\includegraphics [width=0.45\textwidth]{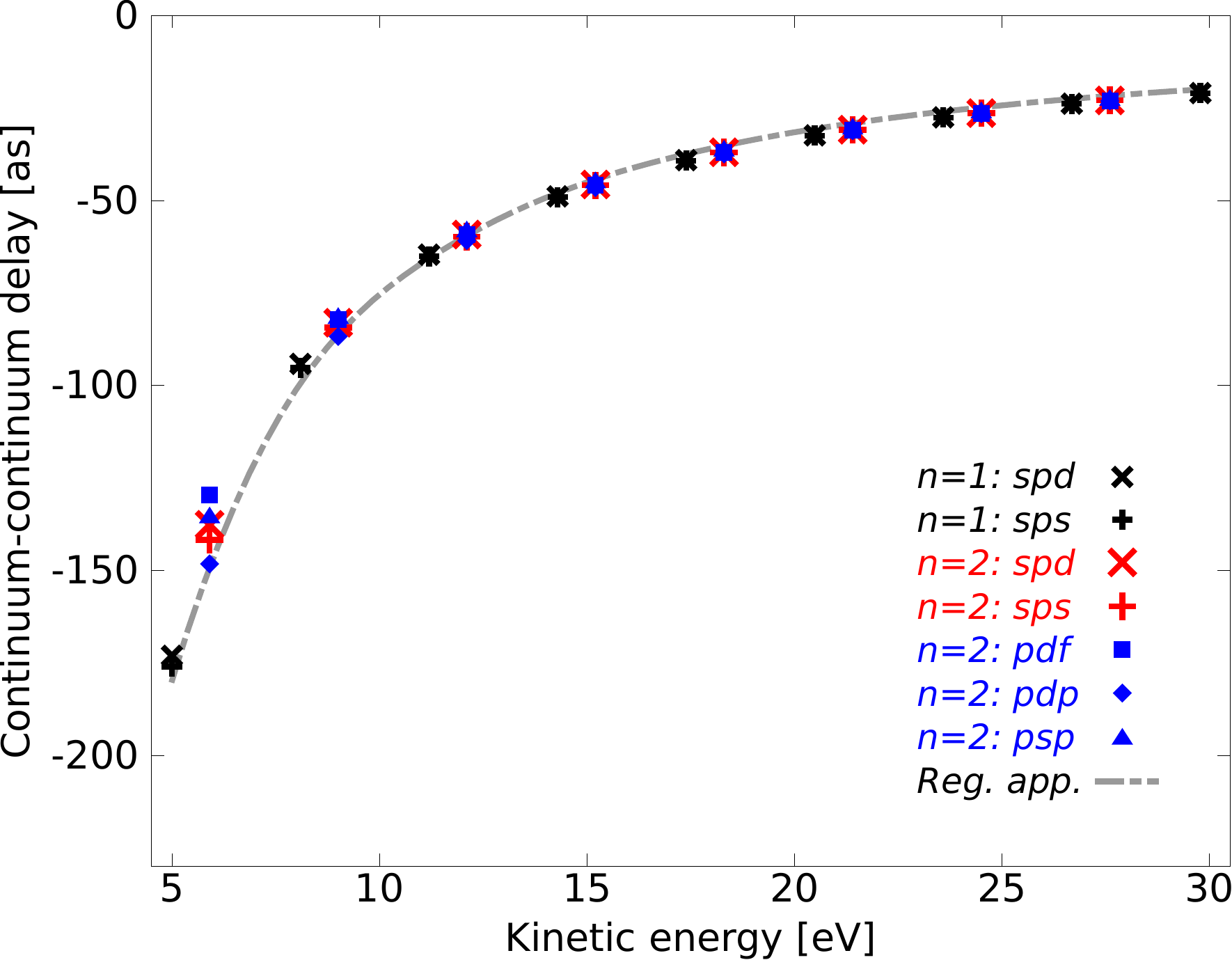}
	\caption{\label{results2}
Continuum--continuum delays from exact calculations in hydrogen from different intial states:  
$1s$ (black symbols),
$2s$ (red symbols) and
$2p$ (blue symbols).
This data demonstrates the {\it universality} of the continuum--continuum delay 
and that it is valid not only for initial states of $s-$character.
The exact data is in excellent agreement with the regularized approximation.
The data correspond to $Z=1$ and a laser probe with $\omega=1.55$~eV.
	}
\end{figure}

In Fig.~\ref{results2}, we address the question of the  universality of $\tau_{cc}$ 
by computing the exact two-photon phases from three different intial states in hydrogen: $1s$, $2s$ and $2p$.
We find that all seven different angular-momentum sequences, $\ell_i\rightarrow \lambda\rightarrow L$, line up on the same universal curve, 
in excellent agreement with the regularized continuum--continuum delay.
In this way, we have verified that not only $s-$type initial states have similar $\tau_{cc}$,
but also initial $p-$states with non-zero angular momentum.

\begin{figure}[h]
	\centering
		\includegraphics [width=0.45\textwidth]{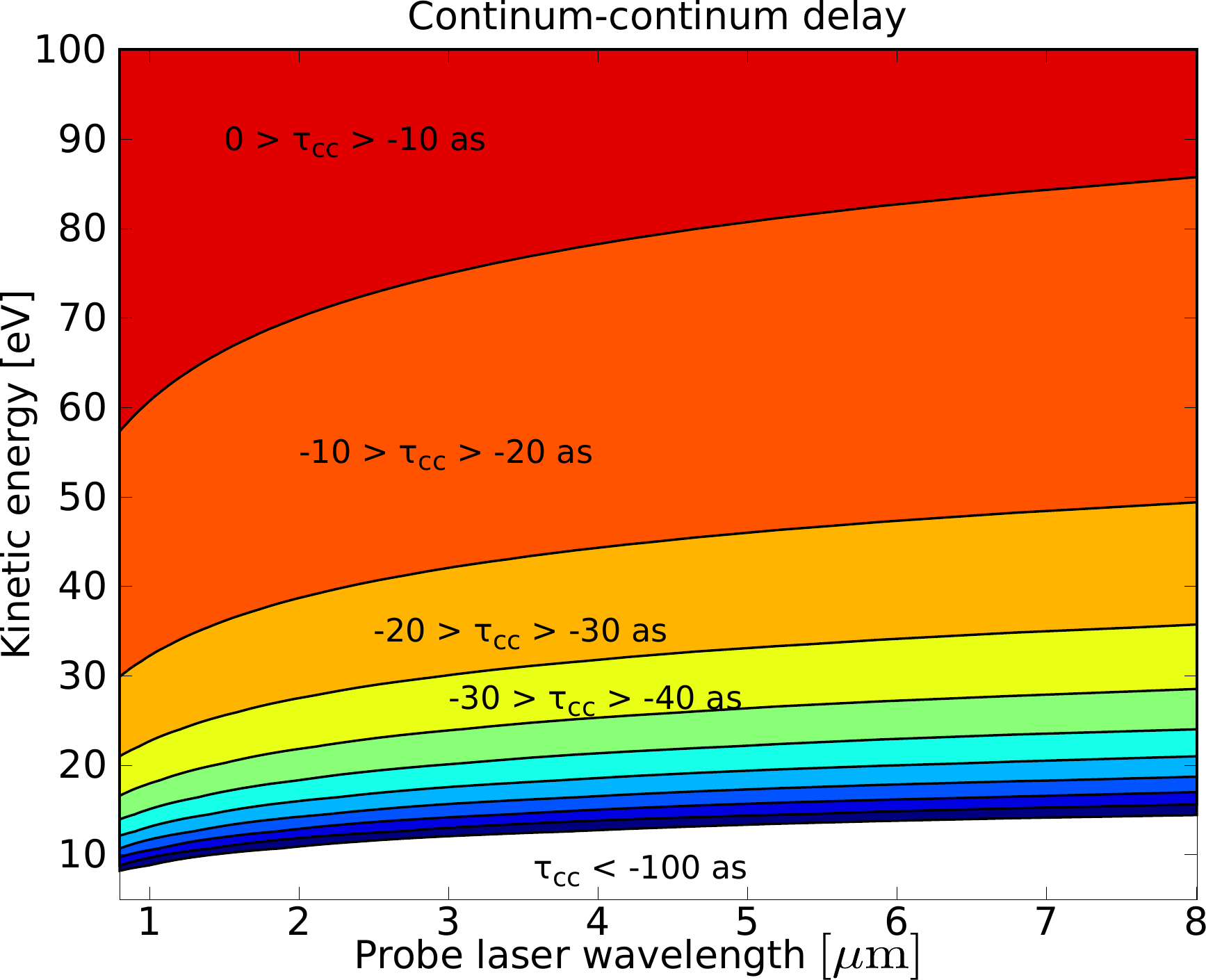}
	\caption{\label{results3a}
Contour plot of continuum--continuum delays at different kinetic energy and laser probe wavelength.
The delays are calculated using the regularized asymptotic approximation.
	}
\end{figure}

\begin{figure}[h]
	\centering
		\includegraphics [width=0.45\textwidth]{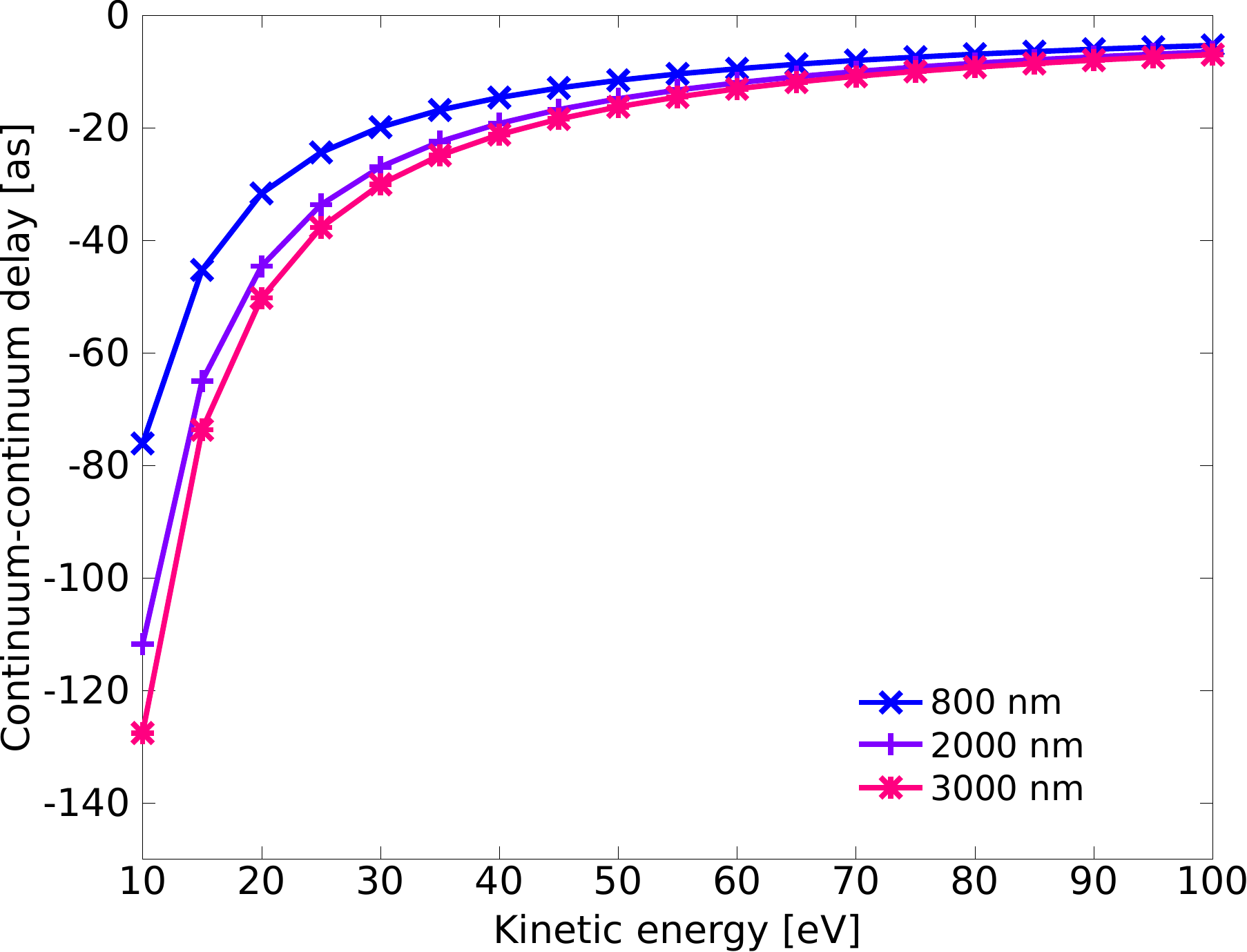}
	\caption{\label{results3b}
Continuum--continuum delays for different IR laser photon energies: 
1.55~eV ($\times$ symbol),
0.62~eV ($+$ symbol) and
0.41~eV ($*$ symbol). 
The delays are calculated using the regularized asymptotic approximation.
	}
\end{figure}

In Fig.~\ref{results3a}, we present a contour plot that provides a rough overview of the magnitude of 
$\tau_{cc}$  as a function of the kinetic energy of the photoelectron and of the wavelength of the probe field. 
In general, $\tau_{cc}$  decreases with the kinetic energy 
and it increases with the wavelength of the probe field.
We note that a softer probe photon leads to an increased delay. 
A few selected $\tau_{cc}$  curves are displayed in Fig.~\ref{results3b}
for a more quantitative comparison at some experimentally relevant wavelengths.
We conclude that $\tau_{cc}$ is not extremely sensitive to the probe wavelength,
except at low kinetic energies, where it takes larger negative values when the probe wavelength increases.
As already mentioned, the delays converge to zero as the kinetic energy is increased, 
however, the convergence is rather slow and there is still $\sim -10$~as of delay remaining 
at a relatively high kinetic energy of $\sim100$~eV.  

\begin{figure}[h]
	\centering
		\includegraphics [width=0.45\textwidth]{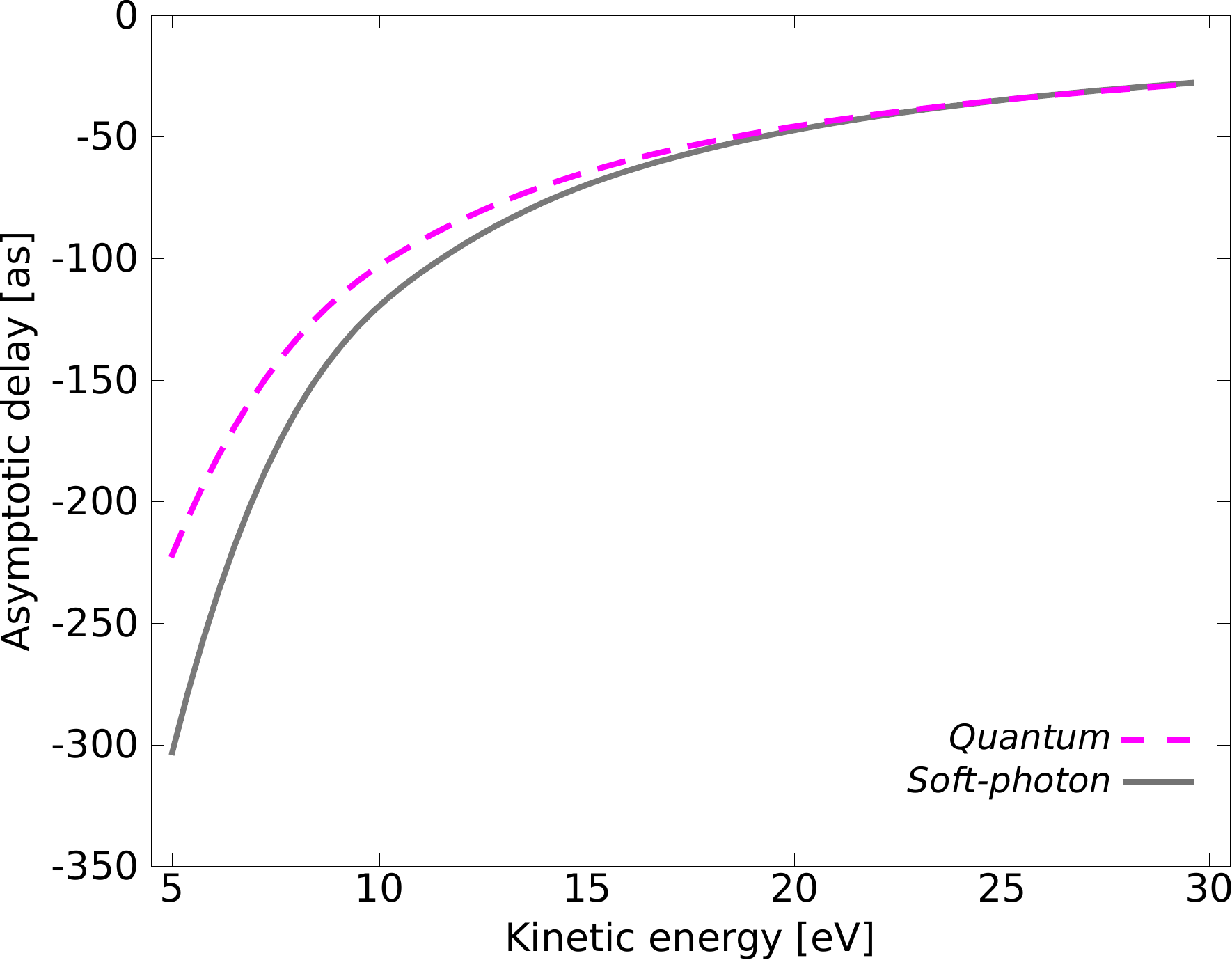}
	\caption{\label{results4}
Continuum--continuum delays calculated using the asymptotic approximation (dashed curve) using Eq.~(\ref{phicc}). They are compared to the corresponding soft-photon limit (curve) using Eq.~(\ref{phiccsoft}). The data correspond to $Z=1$ and to a laser probe with $\omega=1.55$~eV.
	}
\end{figure}

Finally, in Fig.~\ref{results4}, we examine the validity of the soft-photon approximation by comparing 
 $\tau_{cc}$ and $\tau_{cc}^{(soft)}$, calculated 
using Eq.~(\ref{phicc}) and Eq.~(\ref{phiccsoft}), respectively.
The soft-photon limit over-estimates the magnitude of the delay at low energies, 
but it converges rapidly towards the quantum mechanical result as the kinetic energy is increased.

\subsection{Conclusions}
 

The main result of this work is the determination of the {\it phase} associated with 
two-photon transition matrix elements for laser-assisted XUV photoionization, namely Eq.~(\ref{phiMas1}). 
These phases have broad applications in attosecond science 
as they enter naturally in most characterization methods, 
as well as in quantum control schemes and delay experiments of more general character. 
We have demonstrated that the phase is composed of two distinct atomic contributions: 
i) The one-photon scattering phase of the intermediate state  and ii) a quantity, given in Eq.~(\ref{phicc}), that we call the continuum--continuum phase. 
The latter is {\it universal} and 
it describes the added phase induced by the transition from the intermediate continuum state 
to the final continuum state. 
It is independent of the short-range behavior of the atomic potential 
and it depends only on three quantities: the final momentum, the laser-probe frequency and the charge of the remaining ionic core. 
%
%
As expected from the strong-field approach, 
the transition phase is reduced 
to that of the one-photon (intermediate) scattering phase
when the charge of the remaining ion is neglected, 
{\it i.e.} for a short-range potential. 
%
Other interesting findings are related to the classical--quantum correspondence: 
First, the phase of the classical dipole, for exchange of radiation in the continuum, Eq.~(\ref{C_phase}), is closely related, but not identical, to the soft-photon limit of the quantum mechanical continuum--continuum phase, Eq.~(\ref{phiccsoft}).
Further, as epitomized in Eq.~(\ref{r2C}), there exists a  relationship between the initial radial position of the ejected electron (a classical concept) and the asymptotic quantum phase.
We have also demonstrated how the theory of complex transition matrix elements, originally developed for RABBIT, can help to better understand attosecond streaking measurements. 
In our interferometric interpretation of streaking, we stressed the benefits of a weak and monochromatic laser probe field.  Given such experimental conditions, quantitative analysis of more complex systems can be carried out using two-photon transition matrix elements, corrected by many-body perturbation theory.


\subsection*{Acknowledgements}
This research was supported by the Marie Curie program ATTOFEL (ITN), the European Research Council (ALMA), the Joint Research Programme ALADIN of Laserlab-Europe II, the Swedish Foundation for Strategic Research, the Swedish Research Council, the Knut and Alice Wallenberg Foundation, the French ANR-09-BLAN-0031-01 ATTO-WAVE program, COST Action CM0702 (CUSPFEL). 

Parts of the computations have been performed at IDRIS, and part of the work was carried out at the NORDITA workshop: ``Studying Quantum Mechanics in the Time Domain'', in Stockholm. 
One of us (J.M.D.) would like to thank Professor Eva Lindroth for helpful discussions.

\section*{Appendix A: Formal derivation of the approximate expression for the radial component of the 2-photon ATI transition amplitudes}

The radial component of the two-photon ATI matrix elements given in Eq.~(\ref{Tell}) can be rewritten:
\begin{align}
T_{L,\lambda,\ell_i}(k; \epsilon_i +\Omega)   = \lim_{\varepsilon \to 0^+} \langle R_{k ,L} | r\: G_\lambda (\epsilon_i +\Omega+i\varepsilon)\: r |R_{n_i,\ell_i}\rangle.
\label{TGreen}
\end{align}
Here the initial atomic state is represented by the radial wavefunction $R_{n_i,\ell_i} (r)$, while $ R_{k ,L} (r) $ is the wavefunction of the photoelectron with energy $k^2 /2 = \epsilon_i +\Omega \pm \omega$. $G_\lambda (r_2, r_1;\epsilon_i +\Omega)$ is the radial component, of the Green's function given in Eq.~(\ref{Green}) for angular momentum $\lambda$ and positive energy argument $\kappa^2 /2=\epsilon_i +\Omega > 0$. We note the presence of the positive imaginary infinitesimal $ i\varepsilon$ in the argument of $G$: It corresponds to the change $\kappa \to \kappa + i \varepsilon/\kappa$, which  ensures the presence of a converging factor  $e^{-\epsilon r}$ in the integrals below, which otherwise would be divergent. For the sake of conciseness, we will omit this factor in the following.

In the limit of large values of the coordinates $r_1$ and $r_2$, the Green's function takes the following limiting form, written for the case of a Coulomb potential, \cite{Zon, Edwards}:
\begin{eqnarray}
G_\lambda (r_2, r_1;\epsilon_i +\Omega) \approx -{2\over r_1 r_2 \kappa} e^{i[\kappa r_> + \Phi_{\kappa,\lambda} (r_> )]}\nonumber \\  \times \sin [ \kappa r_< + \Phi_{\kappa,\lambda} (r_< )]
\end{eqnarray}
where $r_>$ (resp. $r_<$) denotes the larger (resp. smaller) of $(r_1, r_2)$ and the phases $\Phi_{\kappa,\lambda} (r )$ are defined in Eq. (\ref{Phase}).

We consider now the case of two-photon ATI from the $1s$ state of a hydrogenic system with $R_{1,0}(r) = C_{1,0}  e^{-Zr}$ and normalization constant $ C_{1,0} = 2 Z^{3/2}$, while the photoelectron is described by a continuum wavefunction with asymptotic form given in Eq. (\ref{RkL}).
In the case considered here, $\lambda =1$ while  $L = 0,2$.
Replacing in the general expression of the amplitude Eq. (\ref{TGreen}), one has:
\begin{align}
T_{L,1,0}(k; \epsilon_{1s} +\Omega) = -{2\over k}C_{1,0} N_k \int_0^\infty \!\! \!dr_2 r_2  \sin[k r_2+\Phi_{k,L}(r_2)]\nonumber \\   \int_0^\infty\!\! \!dr_1 r_1^2 e^{-Zr_1}  e^{i[\kappa r_> + \Phi_{\kappa,\lambda} (r_> )]} \sin [ \kappa r_< + \Phi_{\kappa,\lambda} (r_< )]
\end{align}
and splitting the integration ranges over $r_>$ and $r_<$, one gets: 
\begin{align}
T_{L,1,0}(k; \epsilon_{1s} +\Omega)  = -{2\over k}C_{1,0} N_k  \int_0^\infty \!\! \!dr_2 r_2 \sin[k r_2+\Phi_{k,L}(r_2)]\nonumber \\ \times \{ e^{i[\kappa r_2+\Phi_{\kappa,1}(r_2)]}  \int_0^{r_2}\!\! \!dr_1 r_1^2 e^{-Zr_1}  \sin [\kappa r_1 + \Phi_{\kappa,1} (r_1)] \nonumber \\
+  \sin [\kappa r_2 + \Phi_{\kappa,1} (r_2)] \int_{r_2}^\infty \!\! \!dr_1 r_1^2 e^{-Zr_1} e^{i[\kappa r_1 + \Phi_{\kappa,1} (r_1)]} \} 
\end{align}
From the two $r_1-$integrals present within the braces, the first one is by far dominant. This comes from the presence of the exponentially decaying wavefunction of the ground state, which ensures that the integration range containing the origin is dominant. Thus, one can neglect the second term and one has:
\begin{align}
T_{L,1,0}(k; \epsilon_{1s} +\Omega)  \approx - {2\over k}C_{1,0} N_k \nonumber \\ \int_0^\infty \!\! \!dr_2 r_2  \sin [k r_2+\Phi_{k,L}(r_2)]  e^{i[\kappa r_2+\Phi_{\kappa,1}(r_2)]} \nonumber \\ \times \int_0^{r_2}\!\! \!dr_1 r_1^2 e^{-Zr_1}  \sin [\kappa r_1 + \Phi_{\kappa,1} (r_1)]
\end{align}
We remark that the $r_1$-integral contained in this expression is real: $\int dr_1 ... \in \mathbf{R}$ and its precise value does not affect the overall phase of the amplitude which becomes: 
\begin{align}
\arg [T_{L,1,0}(k; \epsilon_{1s} +\Omega)] \approx \nonumber \\ \arg \{ -  \int_0^\infty \!\! \!dr r \sin [k r +\Phi_{k,L}(r)]  e^{i[\kappa r + \Phi_{\kappa,1} (r)]}\}
\end{align}
or, making explicit  the expressions of the Coulomb phases $\Phi$, writing the sine function under its exponential form and keeping only the $r-$dependent terms in the integrand, the amplitude can be expressed in terms of the integrals $J_\pm$ defined in Eq. (\ref{Jint}). Following the same line of reasoning as the one followed in Sec \ref{sec:asympapp}, one can neglect the contribution of the $J_+$ integral, so that:  
\begin{align}
\arg [T_{L,1,0}(k; \epsilon_{1s} +\Omega)] \approx \arg \{ (-)^{{L+2\over2}} (2\kappa)^{iZ/\kappa} e^{i\sigma_1 (\kappa)}\nonumber \\   (2k)^{-iZ/k} e^{-i\sigma_L (k)} \int_0^\infty \!\! \!dr r^{1 +iZ(1/\kappa - 1/k)} e^{i(\kappa -k ) r} ] \} .
\end{align}
where the last integral represents a Gamma function of complex argument times  algebraic factors, so that one recovers the expression of the phase of the radial amplitude given in Eq. (\ref{argTas1}).

This expression, which has been established for an initial $1s$ hydrogenic state with nuclear charge $Z$, remains valid for any $s$ bound state with an exponentially decaying wavefunction. As already mentioned, for non-hydrogenic systems, the Coulomb phase-shifts have to be replaced by the relevant scattering phase-shifts . 

For the sake of reference, we give the complete expression of the asymptotic form of the amplitude, including real factors:
\begin{align}
T_{L,1,0}(k; \epsilon_{1s} +\Omega) \approx (-)^{{L+2 \over2}} {1\over k}C_{1,0} N_{k } e^{-{\pi Z\over2}({1\over \kappa} -{1\over k}) } \nonumber \\  {(2\kappa)^{iZ/\kappa}\over (2k)^{iZ/k} } e^{i[\sigma_1 (\kappa) -\sigma_L (k)]}  {\Gamma[2 +iZ(1/\kappa - 1/k)] \over (\kappa-k)^{2 +iZ(1/\kappa - 1/k)}}\nonumber \\ \times  \int_0^\infty \!\! \!dr r^2 e^{-Zr}  \sin [\kappa r + \Phi_{\kappa,1} (r)] 
\end{align}

\bibliographystyle{elsarticle-num}

\end{document}